\documentclass{elsart}

\usepackage{epsfig}

\newcommand{\MAC}[0]{{\sc mac}}
\newcommand{\FIFPO}[0]{{\sc fifpo}}

\newcommand{\figref}[1]{Fig.\ \ref{#1}}
\newcommand{\eqref}[1]{Eq.\ (\ref{#1})}

\begin{document}

\begin{frontmatter}

\title{Impact of network structure \\
       on the capacity of \\
       wireless multihop ad hoc communication}

\author[label1,label2]{Wolfram Krause}
  \ead{krause@th.physik.uni-frankfurt.de}
\author[label1,label3]{Ingmar Glauche}
  \ead{glauche@theory.phy.tu-dresden.de}
\author[label1]{Rudolf Sollacher}
  \ead{rudolf.sollacher@siemens.com}
\author[label1]{Martin Greiner}
  \ead{martin.greiner@siemens.com}
\address[label1]{Corporate Technology, Information \& Communications, 
                 Siemens AG, 
                 D-81730 M\"unchen, Germany}
\address[label2]{Institut f\"ur Theoretische Physik,
                 Johann Wolfgang Goethe-Universit\"at,
                 Postfach 11 19 32,
                 D-60054 Frankfurt am Main, Germany}
\address[label3]{Institut f\"ur Theoretische Physik,
                 Technische Universit\"at,
                 D-01062 Dresden, Germany}

\begin{abstract}
As a representative of a complex technological system, so-called
wireless multihop ad hoc communication networks are discussed. 
They represent an infrastructure-less generalization of todays
wireless cellular phone networks. Lacking a central control
authority, the ad hoc nodes have to coordinate themselves such
that the overall network performs in an optimal way.
A performance indicator is the end-to-end throughput capacity.
Various models, generating differing ad hoc network
structure via differing transmission power assignments, are 
constructed and characterized. They serve as input for a generic 
data traffic simulation as well as some semi-analytic estimations.
The latter reveal that due to the most-critical-node effect the 
end-to-end throughput capacity sensitively depends on the 
underlying network structure, resulting in differing scaling laws 
with respect to network size.
\end{abstract}


\end{frontmatter}

\newpage
\section{Introduction}
\label{sec:intro}

Complex networked systems are widespread in nature, society and
engineering. Some examples from biology are regulatory gene and
metabolic networks, the neural network of the brain, the immune system,
and, on larger scales, food webs and the ecosystem. Another example, 
now from the social sciences, is multiagent economics, where a 
multitude of large and small traders are interwoven together to curse 
the course of exchange rates and stock prices. Also engineering 
contributes with communication networks, such as the Internet, the 
world wide web and grid computing.

Recently physics has started a new branch, the Statistical Physics
of complex networks \cite{ALB02,DOR03,NEW03}, which takes a generic
and unifying perspective at all of those examples. So far, most of the 
focus has been on the structure of such networks. The analysis of a
great deal of the above mentioned examples has led to the unifying
scale-free discovery and its respective growth modeling with 
preferential attachment. Beyond structure, it is now also dynamics
on and function of networks, which are about to move onto center
stage. The new insight which along these lines has already been given
to regulatory gene networks \cite{ALB03} constitutes for sure a
remarkable highlight. The impact of structure on dynamics is also key 
to epidemic spreading and the proposal of efficient immunization 
strategies for populations and computer networks \cite{COH02}. 

The structure and dynamical properties of engineered communication 
networks in general and computer as well as Internet traffic in 
particular were also discussed heavily within the physics community 
over the last years. Key issues have been besides network structure 
\cite{FAL99,CHE02,CHA03} also phase transition like behavior from a 
noncongested to a congested traffic regime 
\cite{FUK99,FUK01,SOL01,VAL02} and  selfsimilar data traffic 
\cite{PAR00}. So far the analysis of network structure and dynamics 
has been mostly separated from each other. Only very recently a first
coupling of these issues has been picked up, focusing on the impact of
structure on synchronization dynamics \cite{TOR03}. 
--
In this Paper we introduce a new and intriguing complex technological
system to the Physics community, so-called wireless multihop ad hoc
communication networks \cite{MANET,NISTB,MOB02,MOB03}, and 
discuss the impact of network structure onto their performance to 
handle data traffic.

Wireless multihop ad hoc communication networks represent an
infrastructureless generalization of todays wireless cellular phone 
networks \cite{PRO95,PRA98,RAP99}. Lacking a central control authority 
in the form of base stations, each end device acts as router and relays 
packets for other participants. End-to-end communications are possible 
via multihop connections. In order to ensure network connectivity,
efficient discovery and execution of end-to-end routes and avoidance of 
data packet collisions on shared radio channels, the participating 
devices need coordination amongst themselves. Proposals
\cite{ROD99,GLA03} for such a coordination have already been put 
forward upon focusing on the connectivity issue. For the other two 
issues, routing and medium access control, the coordination is much 
more delicate. Routing efficiency would require rather short 
end-to-end routes, resulting in a small network diameter. Due to the 
overall presence of interference, medium access control takes care of 
collision avoidance for data packets traveling on different routes. It 
blocks all neighboring devices of an active one-hop transmission and 
thus would rather favor a small neighborhood, implying a small node 
degree. The opposite demands of these counteracting mechanisms leave 
the network in a state of frustration. This opens the door for the 
generic Statistical Physics of complex networks, asking the question 
what is the impact of network structure on the performance of wireless 
multihop ad hoc communication.

Well-performing network structures would be those which find an
efficient compromise out of this frustration. They would come with a
small end-to-end time delay and a large end-to-end throughput. In 
this Paper we analyze and compare different wireless multihop ad hoc 
network structures that are created by different transmission power 
assignments. Note that the regulation of its transmission power is 
the only intrinsic control action a communication node is able to 
perform to enforce a certain structure of the network. A generic data 
traffic simulation as well as a semi-analytic approach is used to 
calculate the end-to-end throughput associated to these 
network structures. This allows us to study the impact 
of network structure on the data-traffic performance of such 
communication networks.

Sect.\ 2 explains the used network model and various transmission
power assignments. Some properties of the resulting geometric network
graphs are also discussed. The performance of these network structure 
classes with respect to end-to-end throughput is evaluated in Sect.\ 
\ref{sec:traffic} by employing a generic data traffic simulation, the 
emphasis being on scalability. Semi-analytic insights are presented 
in Sect.\ \ref{sec:meanfield} and finally a conclusion is given in 
Sect.\ \ref{sec:conclusion}. The Appendix summarizes the 
graph-theoretical variables, which are used throughout the Paper.

\section{Network structure models for wireless multihop ad hoc
communication}
\label{sec:prelim}

In this Section we are only concerned with structural aspects of 
wireless multihop ad hoc communication networks. Upon invoking some 
justifiable idealizations in Subsect.\ \ref{subsec:networkmodel}, 
various classes of random geometric graphs are constructed in 
Subsects.\ \ref{subsec:constp}-\ref{subsec:shortestvsenergy}, 
which we will from then on reference as network models I-V. The 
various network models differ in the way how transmission power is 
assigned to the communication nodes. Their structural properties will 
be discussed in Subsect.\ \ref{subsec:properties}.

\subsection{Network models in general}
\label{subsec:networkmodel}

Wireless multihop ad hoc networks consist of communication nodes, which
are distributed in space and communicate to each other via a wireless 
medium. For simplicity, mobility of the communication nodes is 
discarded. Exactly $N$ communication nodes are given random positions 
$(x,y) \in [0,L] \times [0,L]$ confined to a square area. Other spatial 
point patterns, e.g.\ of clustered multifractal or Manhattan type, have 
already been discussed in connection with the connectivity issue 
\cite{GLA03}, but will not be given further consideration here.

The radio propagation medium and the receiver characteristics determine
whether nodes can communicate to each other. According to a simple
pro\-pa\-ga\-tion-receiver model, a node $i$ is able to listen to a 
transmitting node $j$, if relative to a $noise$ the power received at 
node $i$, is larger than the signal-to-noise ratio $snr$:
\begin{equation}
\label{eq:propagation}
  \frac{P_j/r_{ij}^\alpha}{noise} \geq snr
  \; .
\end{equation}
$P_j$ denotes the transmission power of node $j$ and $r_{ij}$
represents the distance between the two nodes. Without any loss of
generality we choose the normalization $noise \cdot snr =
1/(\sqrt{2}L)^\alpha$, so that $P=1$ for $r=\sqrt{2}L$.  Depending
on specific in-/outdoor propagation, the path-loss exponent typically
falls into the regime $2\leq\alpha\leq 6$ and is assumed to be
constant. A specific fixation of its value is not required for our
generic view advocated in this Paper. Once (\ref{eq:propagation}) is 
fulfilled, it defines a one-directed communication link 
$j\rightarrow i$. If also the link $i\rightarrow j$ exists, then the 
two nodes are connected via a bidirectional link $i\leftrightarrow j$
and are able to directly communicate back and forth to each other. 
Although not strictly required, bidirectional links are preferred for 
the operation of wireless ad hoc networks, because many protocols 
require instant feedback. Note also, that according to
(\ref{eq:propagation}) the communication links are defined in the
traffic-free regime. This is an additional simplification and does not 
take interference into account. 

En route to the construction of random geometric ad hoc graphs one
further step has to be taken. It deals with power assignment and gives
a specific transmission power value to each node. According to
(\ref{eq:propagation}) this value can be translated into a
transmission range. Those nodes which then fall inside this
transmission range are then the neighbors, to which the picked node is
able to communicate via outgoing wireless links. Power assignment can 
be done via a number of different approaches. Some of them are 
presented in the following Subsections and lead to network models I-V.
The construction of some of these models is based on standard as well 
as non-standard graph-theoretical variables, which for convenience are 
all listed in the Appendix.

\subsection{Network model I: constant transmission power}
\label{subsec:constp}

\begin{figure}
\begin{centering}
\noindent
\epsfig{file=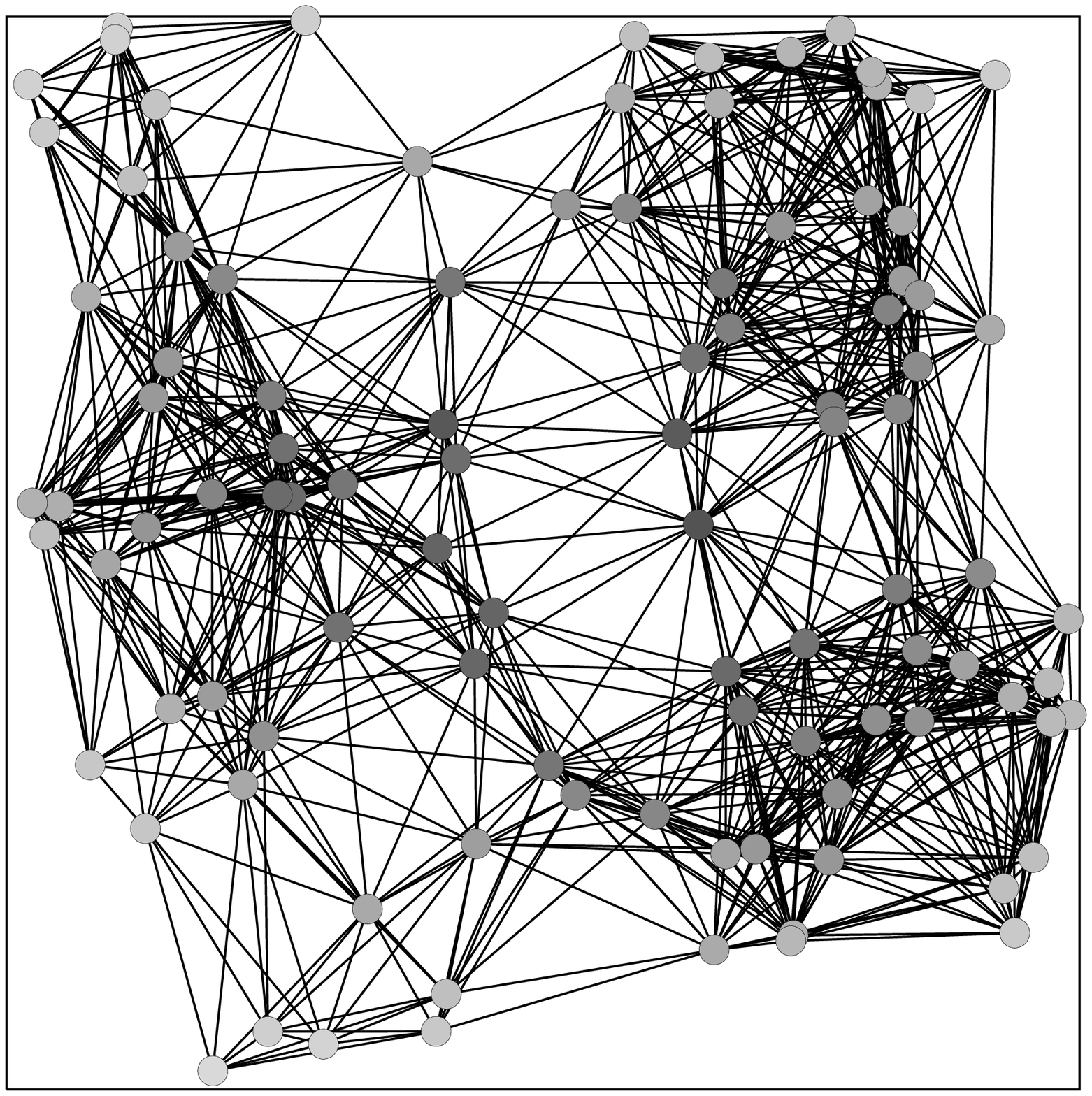,width=6.5cm}
\epsfig{file=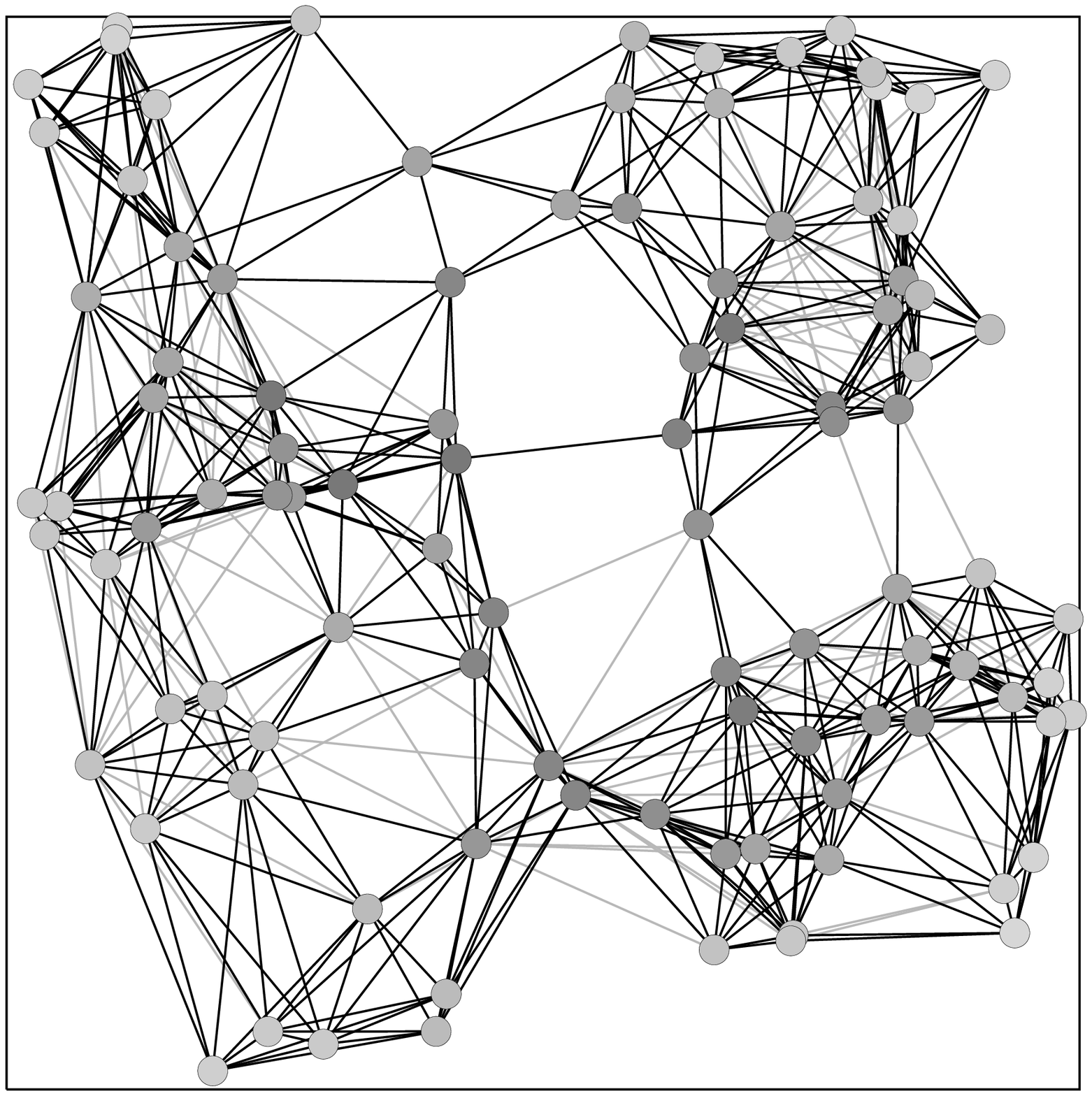,width=6.5cm}
\epsfig{file=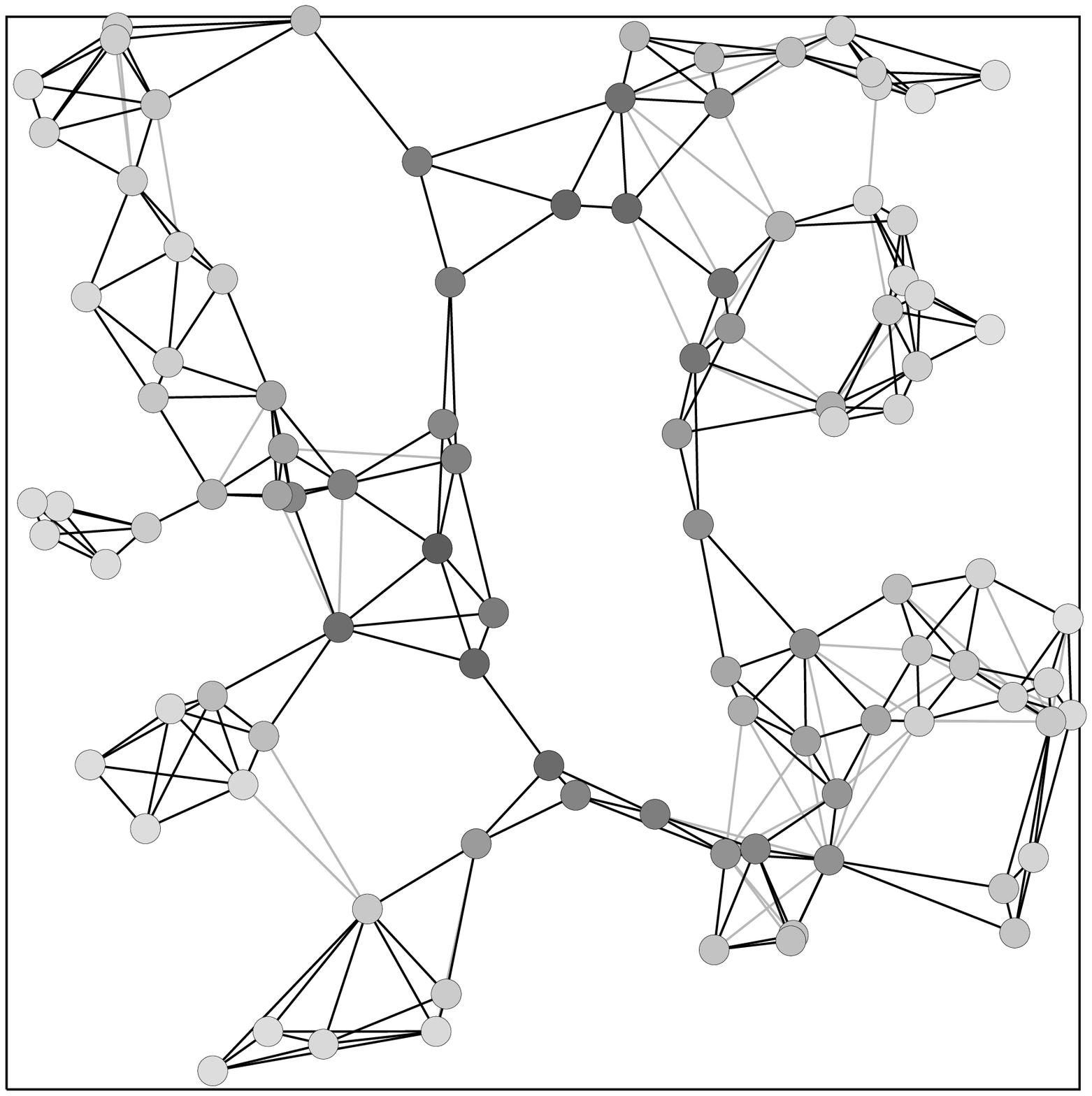,width=6.5cm}
\epsfig{file=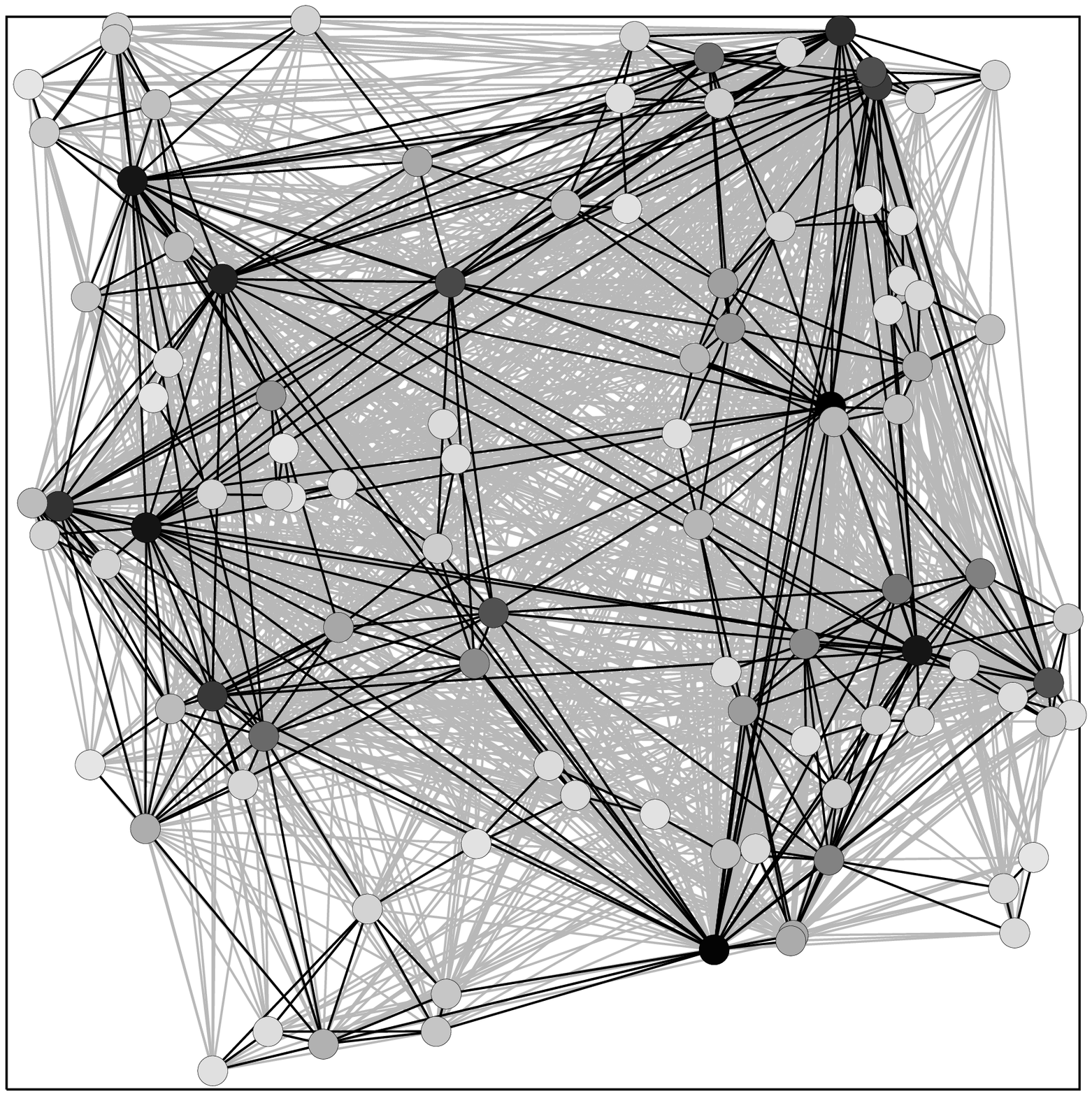,width=6.5cm}
\epsfig{file=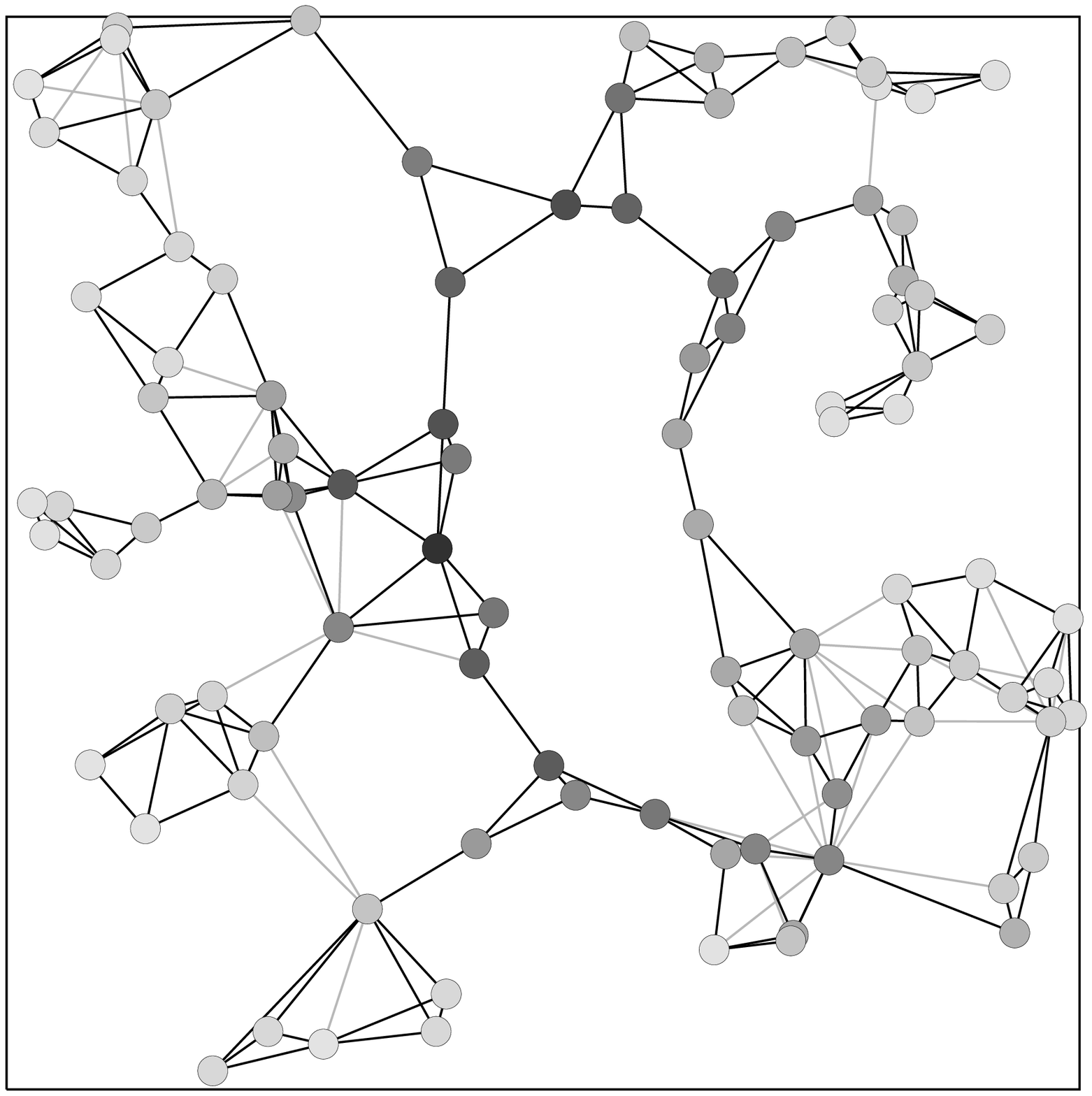,width=6.5cm}
\epsfig{file=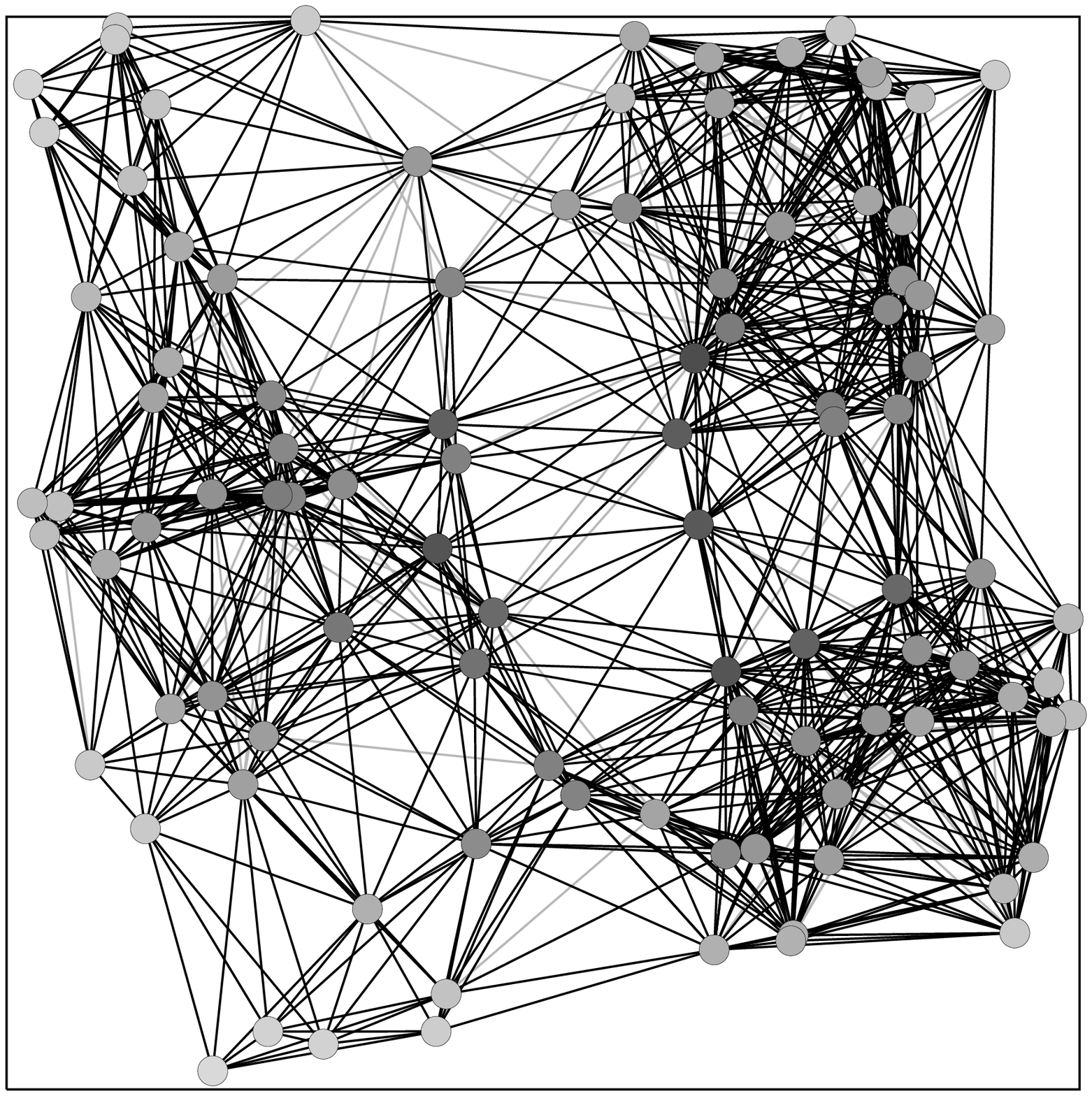,width=6.5cm}
\caption{
Wireless multihop ad hoc graphs based on random homogeneous point 
patterns with $N=100$ nodes: 
(upper left) network model I with $\langle k_i\rangle_\infty=24$,
(upper right) network model II with $k_{min}=8$,
(middle left) network model III with $k_{target}=5$,
(middle right) network topology IV with $\gamma = 2.1$, 
network model V with (lower left) $\lambda=0.0$ and
(lower right) $\lambda=0.8$.
Bi-/onedirected links are shown in black/gray. The gray scale attached
to the nodes reflects their cumulative node inbetweeness $B_i^{cum}$.
} 
\label{fig:networks}
\end{centering}
\end{figure}

The simplest structural design of wireless ad hoc networks is to use 
the same transmission power $P$ for each node \cite{GUP00}. All 
existing links are then bidirected. However, the specific choice for 
$P$ needs to be addressed. If $P$ is chosen too small, the network is 
not strongly connected. If $P$ is too large, too much bandwidth is 
given away with an increased blocking by medium access control. Hence, 
in Ref.\ \cite{GUP00} it was suggested to tune $P$ such that the 
network remains just about strongly connected; see also 
\cite{GLA03,BET02,DOU02,XUE03}. Unfortunately this choice for $P$ 
depends on the network size. To proceed further, it is of advantage to
first translate $P$ into a transmission range $R$, which then 
determines the mean node degree $\langle k_i \rangle_\infty$. With 
$P=(R/\sqrt{2}L)^\alpha$ from (\ref{eq:propagation}) and 
$\rho\pi R^2=\langle k_i \rangle_\infty$ with the node density 
$\rho=N/L^2$ we get the relationship 
$P=(\langle k_i \rangle_\infty / 2\pi\rho L^2)^{\alpha/2}$. Note that 
in the limit $N{\to}\infty$ boundary effects can be neglected, so that
$\langle k_i \rangle_\infty = \langle k_i \rangle$. A quick simulation
reveals that strong connectivity is guaranteed almost surely for 
$\langle k_i \rangle_\infty > a + b \log N$; see also Refs.\ 
\cite{BET02,XUE03}. The values $a=12.7$ and $b=2.11$ have been 
determined such that 99\% of $1000$ sampled random network graphs per
network size $N$ had been strongly connected. Since the network size is 
not a control parameter, but limited in view of practical applications, 
we choose $\langle k_i \rangle_\infty = 24$, unless otherwise noted. 
This value guarantees strong connectivity for network sizes up to 
several thousand nodes almost surely. The upper left part of 
\figref{fig:networks} illustrates a connected wireless multihop ad hoc 
graph with this parameter setting.

\subsection{Network model II: minimum node degree}
\label{subsec:mld}

The const-$P$ assignment of the previous Subsection has notorious 
practical drawbacks. First of all, it is a global rule, where nodes 
need to access information about the complete network state. Second, 
the used constant transmission power strongly depends on the nature 
of the spatial point patterns. As demonstrated in Ref.\ \cite{GLA03}, 
for random patterns of clustered multifractal or Manhattan type the 
corresponding $\langle k_i \rangle_\infty$ values are substantially 
larger than the outcome with random homogeneous patterns. Clearly, a 
much more flexible power assignment is needed, which only collects 
local information and acts in a decentralized manner. A promising 
proposal has been given recently and has been tested for the various 
environments: the minimum-node-degree rule \cite{GLA03}.
 
By exchanging so-called ``hello'' and ``hello-reply'' messages each ad
hoc node is able to access direct information from its immediate
neighbors, defined by its links. A simple local observable for a node
is its node degree. Based on this observable alone, the simple
minimum-node-degree strategy for a node is to adjust its transmission
power to have at least $k_{min}$ bidirectional neighbors. While
some of the nodes will end up having exactly $k_{min}$
neighbors, some others are forced to have more than $k_{min}$
neighbors in order to give an additional bidirectional link to a node
which so far has not accumulated enough neighbors. The only parameter,
i.e.\ $k_{min}$, turns out to be remarkably insensitive to the
procedure on how the random spatial point patterns are thrown. For
more algorithmic details and a detailed discussion see \cite{GLA03}.

Again, a simulation proves that once $k_{min}=a+b\log{N}$ with 
$a=3.5$, $b=0.66$, then 99\% or the thrown network graphs are strongly 
connected. The dependency of $k_{min}$ on the network size is much 
smaller than in the case of const-$P$ networks, where $b=2.11$. To be 
on the safe side for all network sizes considered within this Paper we 
fix the minimum-node degree to $k_{min}=8$, unless otherwise noted. The
upper right part of \figref{fig:networks} illustrates a typical 
wireless multihop ad hoc graph based on this parameter setting.

\subsection{Network model III: constant target degree}
\label{subsec:targetdeg}

In the next three Subsections ad hoc network models are again globally 
constructed, respecting special design properties. They serve to
explore new regions of the giant network-state space. For a network
consisting of $N$ nodes basically $(N-1)^N$ different network states
exist. Pick a node and continuously raise its transmission power
starting with zero. At $P_i(1)$, which follows from
(\ref{eq:propagation}), the closest neighbor is able to receive the 
data transmission from $i$, then at $P_i(2)$ the second closest is able 
to listen, and so on, until $P_i(N-1)$ when the most furthest node is 
also happy to understand. Thus, a power ladder with $N-1$ steps can be 
assigned to each node. Having $N$ nodes, this makes $(N-1)^N$ different 
network states. For sure, not all of them qualify because strong 
network connectivity is a necessary prerequisite. Note however, that 
the const-$P$ and minimum-node-degree network state represent just two 
out of these $(N-1)^N$ states, so that there is a good chance to find 
new gold somewhere in this giant network-state space.

The third set of network models is generated by minimizing the 
optimization function
\begin{equation}
\label{eq:zweie1}
  E_{III}
    =  \sum_{i=1}^N \left( k_i - k_{target} \right)^2
       + \epsilon D
       \; .
\end{equation}
In the physics literature such an optimization function is 
traditionally called an energy function. Here it targets all nodes to 
come with the same node degree $k_{target}$. Unless noted otherwise, we 
consider $k_{target}=5$. The second term on the right-hand side of 
(\ref{eq:zweie1}), which goes with the network diameter, is thought of 
as a small ($\epsilon \approx 0$) admixture guaranteeing network 
connectivity; for the case of partitioned networks we set $D=\infty$.

The search for the optimized network state is carried out using
simulated annealing as described for example in Ref.\ \cite{PRE92}.
Starting from a more or less random initial network state, two search
operations are performed in successive time steps: with probability
$p=0.8$ one node is picked at random and its transmission power 
$P_i(n)$ ladder state is changed to $n\pm 1$, or with probability
$1-p$ two nodes are picked at random and their ladder states
$n_i\leftrightarrow n_j$ are exchanged. For each operation the energy 
is evaluated and the change is accepted with a certain probability, 
that depends on a computational temperature. The latter is slowly 
decreased with ongoing search time, so that tentatively regions of the 
network-state space are explored with lower and lower energy. The 
search process stops once a minimum temperature is reached or the 
energy has remained unchanged after a significantly large amount of 
time steps. With the computing facilities available to us,  we set the 
upper limit of the number of nodes for the optimized networks to 100 
and also the ensemble size to $100$. For the parameter choice 
$k_{target}=5$, a typical optimized wireless multihop ad hoc graph is 
exemplified in the middle left part of \figref{fig:networks}.

\subsection{Network model IV: scale-free target degree}
\label{subsec:sftargetdeg}

As a technological complex system the Internet comes with a scale-free 
network structure \cite{FAL99,CHE02,CHA03}. As the Internet and 
wireless networks are both communication networks, they have many 
parallel, but also distinct features. Nevertheless, curiosity begins to 
ask about the Internet-specific scale-free aspect for wireless multihop 
ad hoc networks.

A rather simple, but straightforward version for the construction of 
scale-free wireless multihop ad hoc networks uses a modification of the 
energy function (\ref{eq:zweie1}), namely
\begin{equation}
\label{eq:zweif1}
  E_{IV}
    =  \sum_{i=1}^N \left( k_i - k^{target}_{i} \right)^2
       + \epsilon D
       \; .
\end{equation}
Independently of the other nodes, each node is randomly assigned its own
target degree $ k^{target}_i$ according to a truncated
scale-free distribution 
\begin{equation}
\label{eq:zweif2}
  p(k_{target})
    = \frac{k_{{target}}^{-\gamma}}
           {\sum_{k_{target}=k_{min}}^{k_{max}}
            k_{{target}}^{-\gamma}}
\end{equation}
with $k_{min} \leq k_{target} \leq k_{max}$. As parameter values
we choose $\gamma=2.1$, $k_{min}=3$ and $k_{max}=\min(30,N)$. A
wireless multihop ad hoc graph, resulting from the optimization of the 
energy function (\ref{eq:zweif1}), is shown in the middle right part of 
\figref{fig:networks}.

Note, that the construction (\ref{eq:zweif1}) relies on bidirected 
links. For such a link to exist, the two attached nodes need to have
suitable transmission power values. With other words, both nodes have
to do something to establish a link, and in doing so, additional
one-directed links emerge unintentionally to other nodes in their
spatial surrounding. This is specific to wireless multihop ad hoc 
graphs and is in contrast to other constructions of geometric 
scale-free graphs \cite{ROZ02,BAR03,HER03,MAN03}.

\subsection{Network model V: minimum power consumption vs.\ shortest path}
\label{subsec:shortestvsenergy}

An obvious design principle for the overall wireless network would be 
to consume as little power as possible. A suitable measure for the 
power consumption of node $i$ is given by the product of its 
transmission power $P_i$ and its node inbetweeness $B_i$, the latter 
representing a measure for the frequency to relay packets within the 
network. Another, different design principle might be to reach the 
intended receiver in as few as possible multihop steps in order to 
reduce the end-to-end time delay. The respective measure would be 
simply proportional to the network diameter $D$. Taken together, both 
design principles lead to the energy function
\begin{eqnarray}
\label{eq:zweig1}
  E_V(\lambda)
    &=&  \frac{(1-\lambda)}{N^{(5-\alpha)/2}} \sum_{i=1}^N B_i P_i
         + \frac{\lambda}{\sqrt{N}} D 
         \; .
\end{eqnarray}
The normalizations $N^{(5-\alpha)/2}$ and $\sqrt{N}$ have been 
introduced for the first and second term, which follow from the 
intuitively expected $D\sim\sqrt{N}$, $B_i\sim N D\sim N^{3/2}$ of 
(\ref{eq:zweib4}) and $P_i\sim N^{-\alpha/2}$ of
(\ref{eq:propagation}). $\lambda=0$ reflects the 
first design principle and will lead to rather sparse network 
structures. The second design principle is represented by $\lambda=1$ 
and will prefer fully connected network structures. Since these two 
design principles come with opposite demands, it is interesting to 
study also values of $\lambda$ falling inbetween zero and one. For the 
parameter settings $\lambda=0.0$ and $0.8$ the lower left and right 
parts of \figref{fig:networks} illustrate two examples for respective 
wireless multihop ad hoc graphs.

\subsection{Structural properties of network models I--V}
\label{subsec:properties}

The mean node degree, as well as means and distributions of all 
following single-node variables, is obtained by sampling first over 
all nodes within one random geometric graph realization and then over 
all generated sample realizations representing the corresponding 
network model. A part of Table \ref{tab:data_100} summarizes the means 
of various node and link degrees for network models I--V with $N=100$. 
For network models I and II the mean values are also given for 
$N=2000$. These values show only little variation with the network size 
and can thus be considered to be close to their asymptotic 
$N{\to}\infty$ limit. The power-minimizing limit $\lambda{\to}0$ of 
network model V produces the smallest numbers. The scale-free network 
IV yields the largest mean excess number of one-directed links. The 
reason for this routes in the uncorrelated assignment of each node's 
target degree, which does not adapt to the local structure of the 
spatial point patterns. For all network models the average link degree 
is about a factor $1.25{-}1.45$ larger than the mean node degree, the 
exception being again network model IV, where this factor amounts to 
almost $2$.

\begin{table}
\begin{tabular}{l|rrrrrr|rr}
& \multicolumn{6}{c|}{$N=100$} & \multicolumn{2}{c}{$N=2000$} \\
& $\mathrm{I}_{24}$ 
& $\mathrm{II}_{8}$ 
& $\mathrm{III}_{5}$ 
& $\mathrm{IV}_{2.1}$ 
& $\mathrm{V}_{0.0}$ 
& $\mathrm{V}_{0.8}$
& $\mathrm{I}_{24}$ 
& $\mathrm{II}_{8}$ 
\\ \hline 
$\langle k_i \rangle$ 
&   $18.6$ &   $10.1$ &     $4.9$ &    $6.4$ &  $ 4.7$ & $19.8$	           
&   $22.7$ &    $9.7$ 
\\
$\langle k_{i\leftrightarrow j} \rangle$ 
&   $26.4$ &   $13.3$ &     $6.2$ &   $12.6$ &  $ 6.2$ & $28.5$	           
&   $32.4$ &   $12.9$ 
\\
$\langle k_{i\leftrightarrow j}^{in} \rangle$ 
&   $26.4$ &   $14.6$ &     $8.0$ &   $18.9$ &  $ 6.7$ & $29.1$	           
&   $32.4$ &   $13.8$ 
\\
$\langle k_{i\leftrightarrow j}^{out} \rangle$ 
&   $26.4$ &   $15.3$ &     $8.8$ &   $39.3$ &  $ 7.0$ & $29.2$	           
&   $32.4$ &   $14.1$ 
\\
$\langle C_i \rangle$ 
&   $0.68$ &   $0.64$ &    $0.57$ &   $0.57$ &  $0.54$ & $0.68$	           
&   $0.61$ &   $0.58$ 
\\
$\langle C_{i\leftrightarrow j} \rangle$ 
&   $0.54$ &   $0.50$ &    $0.42$ &   $0.39$ &  $0.41$ & $0.54$	           
&   $0.47$ &   $0.44$ 
\\
$D$ 
&    $2.6$ &    $3.7$ &     $6.4$ &    $4.0$ &  $ 6.3$ & $ 2.4$	           
&   $10.3$ &   $16.4$
\\ 
$\langle B^{cum}_i \rangle/N^2$ 
& $0.55 $ & $0.49 $ & $0.53 $ & $1.16 $ & $0.43 $ & $0.58 $ 
& $0.13$ & $0.10$ 
\\
$\langle \sup B_i \rangle/N^2$ 
& $0.08 $ & $0.16 $ & $0.31 $ & $0.24 $ & $0.23 $ & $0.03 $ 
& $0.04$ & $0.11$ 
\\
$\langle \sup B_{i\leftrightarrow j} \rangle/N^2$ 
& $0.03 $ & $0.09 $ & $0.24 $ & $0.11 $ & $0.30 $ & $0.10 $ 
& $0.02$ & $0.07$ 
\\
$\langle \sup B^{cum}_i \rangle/N^2$ 
& $1.12 $ & $0.97 $ & $1.29 $ & $1.67 $ & $1.14 $ & $1.20 $ 
& $0.35$ & $0.44$ 
\\
$\langle P_i \rangle N^{\alpha/2}$ 
&    $3.8$ &    $2.2$ &     $1.4$ &    $5.3$ &   $0.9$ &  $3.9$		   
&    $3.8$ &    $1.7$ 
\\
$\langle B_i \cdot P_i \rangle$ 
&    $9.8$ &    $8.3$ &    $10.0$ &   $40.1$ &  $ 6.0$ &  $9.7$   
&   $38.8$ &   $30.3$ 
\\
\end{tabular}
\vskip1ex
\caption{
Mean structural properties of wireless ad hoc network graphs resulting 
from models I--V of sizes $N=100$ and $N=2000$.
}
\label{tab:data_100}
\end{table}

Without showing we remark on the node degree distributions. Network 
model I reveals a very broad Gaussian-like distribution. Network model 
II produces a dominant spike at the minimum node degree $k_{min}$, 
which is continued by a small, but broad Gaussian-like tail towards 
larger degrees. As expected, network model III gives rise to a narrow 
distribution centered around $k_{target}$. A scale-free target degree
distribution with $\gamma=2.1$ has been the input of network model IV 
and an approximate scale-free degree distribution is also its output. 
However, the output is modified to $\gamma = 2.5$ (for $N=100$) within 
$5\leq k_i \leq 30$. This is due to the computationally dictated 
smallness of the network, the spatial geometry of the network, the 
node-independence of the degree assignment and the soft square in the 
expression (\ref{eq:zweif1}) of the energy function. The node degree 
distributions of network model V deserve some more attention. The two 
limiting cases are obvious. For $\lambda\to 0$ the transmission-power 
term in (\ref{eq:zweig1}) dominates, leading to a very thin network 
structure with a narrow, small-mean distribution. In the limit 
$\lambda\to 1$ the network-diameter term becomes more and more 
important, leading to an increasingly connected network structure and 
culminating in a fully connected network, once $\lambda=1$. Inbetween 
the two extremes no critical network structure emerges. Due to the 
involved spatial geometry of the node patterns, the resulting 
Gaussian-like node-degree distributions are very robust and lead to 
network structures, which are very similar to the ones obtained with 
network model I with a suitably tuned transmission power $P$.

\begin{figure}
\begin{centering}
\epsfig{file=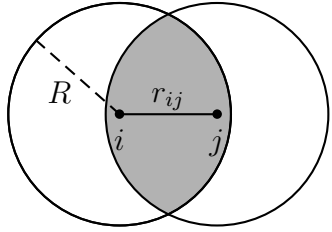,width=4.5cm}
\caption{
Cluster coefficient resulting from network model I: given a node $i$ 
and a neighboring node $j$, the probability that a node $k$ is a 
common neighbor of $i$ and $j$ is given by the quotient of the shaded
area and the area $\pi R^2$ of the transmission disc.  
}
\label{fig:constp_c_analytic}
\end{centering}
\end{figure}

As can be read off from Table \ref{tab:data_100}, all network models
(except model V in the $\lambda\to 1$ limit) produce an $N\to\infty$ 
asymptotic value around $\langle C_i \rangle \approx 0.6$ for the
node-based cluster coefficient. A similar behavior is also observed 
for the link-based cluster coefficient. This observation can be 
explained with the following geometrical argument. For large model I 
networks with constant transmission power, let $R$ be the transmission 
range of all nodes. Pick any node, say $i$, and one of its neighbors, 
say $j$, the distance between the two nodes being $r_{ij}$. The shaded 
area of \figref{fig:constp_c_analytic} describes the locations of 
possible further nodes that have links to both nodes $i$ and $j$. In 
the limit of large $N$, the quotient between this shaded area and 
$\pi R^2$ then describes $j$th contribution to the cluster coefficient 
of node $i$. Using simple geometry, this contribution amounts to
\begin{equation}
\label{eq:dreib1}
  C(r_{ij}) 
    =  \frac{2}{\pi} \left( 
       \arccos \left(\frac{r_{ij}}{2R}\right) 
       - \frac{r_{ij}}{2R} \, \sqrt{1- \left(\frac{r_{ij}}{2R}\right)^2} 
       \right)
       \; .
\end{equation}
The mean cluster coefficient of node $i$ can now be calculated by
averaging $C(r)$ over all $0 \le r \le R$,
\begin{equation}
\label{eq:dreib2}
  \langle C_i \rangle 
    =  \frac{\int_0^R C(r) 2\pi r\d{r}}{\pi R^2} 
    =  1 - \frac{3\sqrt{3}}{4\pi}
    \approx  0.59
       \; .
\end{equation}
This calculation very well explains the observed values around $0.6$, 
which now have to be interpreted as a direct consequence of the 
two-dimensionality of the spatial point patterns. For small networks, 
finite size effects are essential and lead to a larger cluster 
coefficient in almost all cases. 

The network diameter for models I--V is also listed in Table
\ref{tab:data_100}. Models I and II allow to address the dependence of 
$D$ on the network size. For $N>100$ the scaling law $D\sim\sqrt{N}$ is 
found for both types of network models. This outcome is in full 
agreement with our lattice intuition, when $N$ nodes are fully packed 
onto a two-dimensional lattice with side length $\sqrt{N}$. The quality 
of this scaling law degrades for networks of small size. 

The means of node and link inbetweeness are already fixed by the sum
rules (\ref{eq:zweib4}) and (\ref{eq:zweib6}) and are fully expressible 
in terms of the network diameter and the average node degree. As a 
function of the network size and independent of the network type I or 
II, both quantities scale as 
$\langle B_i \rangle \sim \langle B_{i{\leftrightarrow}j} \rangle 
 \sim N^{3/2}$. 
--
It is also interesting to look at the most extreme values. As we will 
realize in Sect.\ \ref{sec:meanfield}, they are of relevance for the 
network performance with respect to data traffic load. Table 
\ref{tab:data_100} also lists the values of 
$\langle \sup_i(B_i) \rangle$ and 
$\langle \sup_{i{\leftrightarrow}j}(B_{i{\leftrightarrow}j}) \rangle$
for network models I--V of size $N=100$ and for network models I and II 
of size $N=2000$, respectively. First the supremum has been taken 
within one network realization and then this supremum has been averaged 
over a large enough sample of realizations belonging to the same 
network model. Table \ref{tab:fit_2000} lists the corresponding scaling 
exponents with respect to the size of network models I and II. The 
exponents are found to deviate noticeably from $3/2$.

\begin{table}
\begin{tabular}{l|rrrrrr}
$\langle\mathcal{O}\rangle\sim N^\gamma$ 
& $\mathrm{I}_{24}$ 
& $\mathrm{II}_{8}$ \\ \hline 
$\langle \sup_i B_i \rangle$ 
&  $1.80$ &  $1.92$ \\
$\langle \sup_{i\leftrightarrow j} B_{i\leftrightarrow j} \rangle$ 
&  $1.90$ &  $1.96$ \\
$\langle B^{cum}_i \rangle$ 
&  $1.49$ &  $1.47$ \\
$\langle \sup_i B^{cum}_i \rangle$ 
&  $1.60$ &  $1.77$ \\
\end{tabular}
\vskip1ex
\caption{
Exponents $\gamma$ of the observed scaling laws 
$\langle\mathcal{O}\rangle \sim N^\gamma$, extracted from network models
I and II of sizes $100\leq N\leq 2000$.
}
\label{tab:fit_2000}
\end{table}

The quantities $\langle B_i^{cum} \rangle$ and 
$\langle \sup_iB_i^{cum} \rangle$ related to the cumulative node 
inbetweeness are also listed in Table \ref{tab:data_100} for the 
various network models. For network models I and II the exponents of 
the observed scaling laws are put down in Table \ref{tab:fit_2000}. 
Whereas $\langle B_i^{cum} \rangle \sim N^\gamma$ comes with the 
expected $\gamma\approx 3/2$, the exponent of 
$\langle \sup_i B_i^{cum} \rangle \sim N^\gamma$ again reveals
a noticeable increase.

By definition the distribution $p(P_i)$ for the transmission power of 
network model I is a $\delta$-function. For network models II, III and 
V $p(P_i)$ is approximately described by a Gamma or log-normal 
distribution. As outlined in \cite{GLA03} this behavior traces back to 
the rather narrow node-degree distributions and the random homogeneity 
of the spatial point patterns. As listed in Table \ref{tab:data_100}, 
the average transmission power $\langle P_i \rangle$ ranks these 
network models in the order I, II, III and V, with model I requiring by 
far the largest value. However, upon switching from 
$\langle P_i \rangle$ to $\langle B_i P_i \rangle$, all models more or 
less come with the same average value. This demonstrates that the power 
consumption of the overall network, where the multihop forwarding of 
end-to-end communications is taken into account, does not sensitively 
depend on the network structure. Of course, the scale-free network 
model IV represents again an exception to this rule.

\section{Generic data traffic and end-to-end throughput}
\label{sec:traffic}

After having presented the various different wireless multihop ad hoc 
network models I--V and having discussed their structural properties, 
we now turn to the dynamics on such networks. A generic data traffic 
simulation will be described in the first Subsection. In the second
Subsection, the end-to-end throughput, which represents the network's 
capacity to handle data traffic without network overloading, will be 
discussed and compared between network models I--V.

\subsection{Generic data traffic simulation}
\label{subsec:traffic}

A detailed, engineering-like approach to data traffic simulation would
explicitly model the exchange of all control messages needed for link, 
medium access and routing control and then would aim to analyze and 
compare different medium access and routing algorithms 
\cite{MOB02,MOB03}. Since for the time being, we are interested in the 
overall behavior of the generated network structures, we prefer to use 
a rather generic approach to data traffic simulation. The simulation 
uses discrete time steps and assumes that all nodes start each time 
step simultaneously. A time step begins with the packet creation phase. 
Then a short contention phase for medium access control is succeeded by 
a longer packet transmission phase.  In the following we give a short 
description of the used generic simulation.

At each time step, a new data packet is created at each node with the
probabilistic packet creation rate $\mu<1$. At a node, its new packet
is assigned a random destination out of the other $N-1$ nodes and is
then put at the end of its buffer queue, assumed to have infinite
capacity. In the future course the packet will be forwarded along the
shortest path to its final destination. The routing matrix 
$\mathcal{R}_{ij}$ takes care of this forwarding, where 
$\mathcal{R}_{ij}=k$ means that a packet at node $j$ with destination 
$i$ has the next hop $k$. The routing matrix has been determined with 
Dijkstra's algorithm \cite{DIJ59}. If shortest-path degeneracy occurs, 
then one of the paths is chosen at random and once and forever kept 
fixed, not only for this specific end-to-end communication, but also 
for all future communications between the same original sender and the 
same final recipient. This procedure is not fully consistent with the 
employed handling of shortest-path degeneracy for the determination of 
the node and link inbetweeness, but on the other hand still very 
similar due to the randomness in the selection process. Nodes, for 
which a new packet has been created, are blocked for the remainder of 
this time step.

During a short contention phase, still at the beginning of the time
step, the remaining non-blocked nodes compete for gaining sender
status. One of them with a nonzero buffer is randomly picked first and
allowed to transmit its first-in-line packet to the envisaged
neighboring node. Both, the sending as well as receiving node block 
their respective one-hop outgoing neighbors for the remainder of this 
time step. This blocking is called medium access control (\MAC) and is 
very necessary to avoid collision of data packets in a wireless medium. 
Then another node from the nonzero buffer and non-blocked list is 
chosen at random and allowed to attempt the transmission of its 
first-in-line packet. If the intended receiver has already been blocked 
before, the node tries to submit its second-in-line packet and so on, 
until either the first idle recipient is found or the end of its buffer 
queue is reached. We denote this strategy as 
first-in-first-possible-out (\FIFPO). If this node succeeds to submit a 
packet, it then (\MAC-) blocks again its remaining outgoing neighbors 
as well as those of the receiving node. This iteration is repeated 
until no free one-hop transmission is left for this time step.

In case of existing one-directed links, which occur for a
heterogeneous assignment of transmission power, it might occur that
a one-directed outgoing link associated to the latest \MAC-operation
blocks a node, which within the contention phase of this time step has
already gained sender or receiver status in a previously approved
one-hop transmission. For such cases, the previously assigned sender
and receiver are blocked again. Furthermore, their outgoing
neighborhoods remain blocked and are not freed.

Then all chosen transmitting nodes submit their selected packet and
remove this packet from their respective buffer list. The receiving
nodes either add their incoming packet to the end of their queuing
list or, if they are the final recipient, destroy this packet. This
concludes the actions taken for this time step and the whole game is
repeated for the next time step.

\subsection{End-to-end throughput}
\label{subsec:trafficresults}

Having generic data traffic simulations now at hand, legitimate
questions to ask are: what is the critical network load, how does it
scale with the network size and how does it depend on the network
structure? The performance of a network depends crucially on the
interplay between routing and medium access control. The adopted
shortest-path routing would prefer network topologies with small
diameter $D$, but then an active one-hop link would come with a large
number $k^{out}_{i{\leftrightarrow}j}$ of attached one-hop neighbors, 
which are \MAC\ blocked. On the other side, if the \MAC\ blocked 
neighborhood is reduced to a minimum, then the network diameter 
increases and it needs more hops to deliver a packet to its final 
destination. Furthermore, not all the nodes are equally loaded. Due to 
their exposed spatial location in the network center some of them have 
to forward more packets than others and come with a larger node 
inbetweeness. The network models I--V, illustrated in 
\figref{fig:networks} and filed in Table \ref{tab:data_100}, all come 
with different weightings of the decisive characteristics like $D$, 
$k^{out}_{i{\leftrightarrow}j}$ and $\sup_i B_i$, so that it is not 
possible to tell from the beginning which of these models performs 
best.

Given a specific network model realization and network size, the 
critical traffic load $\mu_{crit}$ per node is defined as the maximum 
packet creation rate $\mu$, where on average the flux of newly created 
packets is still equal to the flux of end-to-end delivered data 
packets. $\mu < \mu_{crit}$ defines the subcritical phase, whereas
$\mu > \mu_{crit}$ is denoted as the supercritical phase. As each node 
of the network comes with its own in- and out-flux of data packets, the 
critical network load $\mu_{crit}$ is determined by one node, the most
critical node, for which its in- and out-flux 
$\mu^{in}_i = \mu^{out}_i$ become identical. For all the other nodes
their respective in-flux is still smaller than their out-flux, i.e.\
$\mu^{in}_{j\neq i} < \mu^{out}_{j\neq i}$.

The mean in-flux of data packets for node $i$,
\begin{equation}
\label{eq:vierb1}
  \mu^{in}_i
    =  \frac{1}{\left\langle t^{arriv}_i \right\rangle}
       \; ,
\end{equation}
is equal to the reciprocal of the average interarrival time 
$\langle t^{arriv}_i \rangle$, the latter meaning the time lag between 
two successive packet arrivals at node $i$. Two things should be 
remarked here: First, it does not matter if a packet is received via 
transmission from a neighboring node $j\in\mathcal{N}_i$ or if it is 
generated right at node $i$ as long as they are added to the buffer 
queue of $i$. Second, arriving packets to node $i$ with $i$ being the 
final receiver are not counted because they are not handed over to the
buffer queue, but are deleted from the network.

The out-flux of data packets for node $i$,
\begin{equation}
\label{eq:vierb2}
  \mu^{out}_i
    =  \frac{1}{\left\langle t^{send}_i \right\rangle}
    =  \frac{1}{\tau_i}
       \; ,
\end{equation}
is given as the reciprocal of the average sending time; for later
convenience we introduce the abbreviation 
$\langle t^{send}_i \rangle = \tau_i$. Conditioned on a non-empty 
buffer queue, the latter is defined as the time it takes to send the 
next packet from node $i$ to any of its neighboring nodes.  The time 
counter for the sending time is reset either with a sending event from 
an $n_i \geq 2$ buffer queue or with the arrival of a packet to an 
empty $n_i = 0$ buffer queue. 

As function of the packet creation rate the curves for $\mu^{in}_i$
and $\mu^{out}_i$ intersect at $\mu=\mu^{crit}_i$. This intersection
point is different for different nodes. The critical network load is now
given as
\begin{equation}
\label{eq:vierb3}
  \mu_{crit}
    =  \min_i \mu^{crit}_i
       \; .
\end{equation}
This is a per-node quantity. It is related to the overall-network quantity
\begin{equation}
\label{eq:vierb4}
  T_{e2e}
    =  \mu_{crit} N
       \; ,
\end{equation}
which is denoted as the end-to-end throughput. The latter describes the
maximum number of end-to-end communications, which can be completed per
time step without network overloading. In this respect, $T_{e2e}$ can 
also be thought of as a network capacity and $\mu_{crit}$ as a specific
network capacity. 

\begin{figure}
\begin{centering}
\epsfig{file=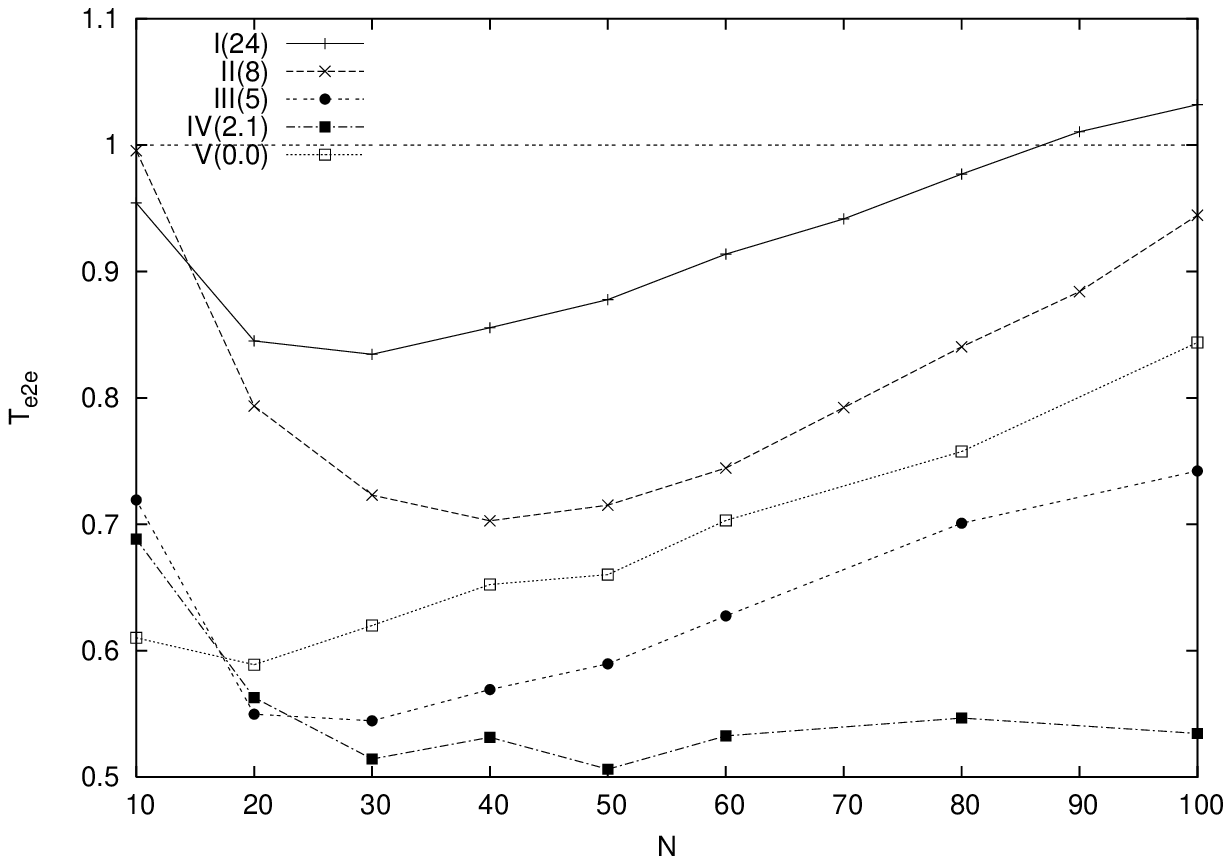,width=12cm}
\epsfig{file=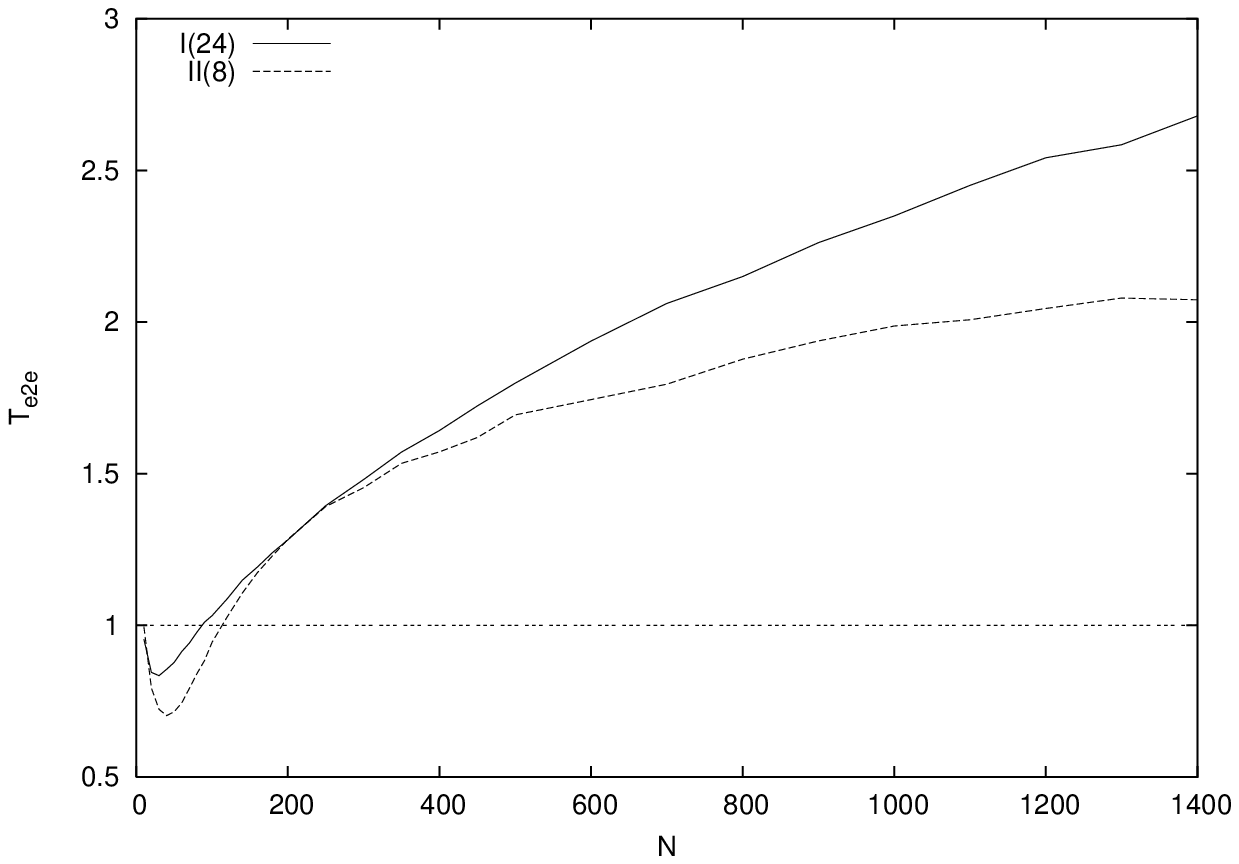,width=12cm}
\caption{
End-to-end throughput as a function of network size 
determined from generic data traffic simulations.
Network models are 
I (vertical crosses) with $\langle k_i \rangle_\infty = 24$, 
II (rotated crosses) with $k_{min}=8$, 
III (full dots) with $k_{target}=5$,
IV (full squares) with $\gamma=2.1$ and
V (open squares) with $\lambda=0.00$.
}
\label{fig:Te2e}
\end{centering}
\end{figure}

\figref{fig:Te2e} summarizes the end-to-end throughput results
obtained from the generic data traffic simulations. For each network
model, $T_{e2e}$ represents an average over a sufficiently large 
ensemble of 100 network realizations. The simulation time of 100000 
time steps for each realization has been large enough for the 
determination of $\mu_{crit}$. For the regime $N \leq 100$ we find that 
network model I with $\langle k_i \rangle_\infty = 24$ yields the 
highest e2e-throughput, followed by network model II with $k_{min}=8$. 
Network model V with $\lambda=0$ comes in at third place and network 
model III with $k_{target}=5$ is fourth. The smallest end-to-end 
throughput is obtained with the scale-free network model IV; a quick 
look again at the middle right part of \figref{fig:networks} reveals 
that the reason for this is the large number of occurring one-directed 
links, which in addition to the bidirected links block large parts of 
the network via medium access control.

The observed ranking of small-sized network models I, II, V and III 
might suggest, that average node degree and end-to-end throughput are
correlated: the larger $\langle k_i \rangle$, the larger $T_{e2e}$. 
This correlation appears to be further supported by a fully connected
network, which comes with the largest possible node degree 
$\langle k_i \rangle = N-1$ and has $T_{e2e} = 1$. Since each node then
has a bidirected link to any other node, medium access control blocks 
the complete network for a single one-hop transmission. This implies 
that at maximum only one packet per time step is delivered to its final 
destination, leading to $T_{e2e} = 1$. For visualization the 
$T_{e2e} = 1$ line is also drawn in Fig.\ \ref{fig:Te2e}.

\begin{figure}
\begin{centering}
\epsfig{file=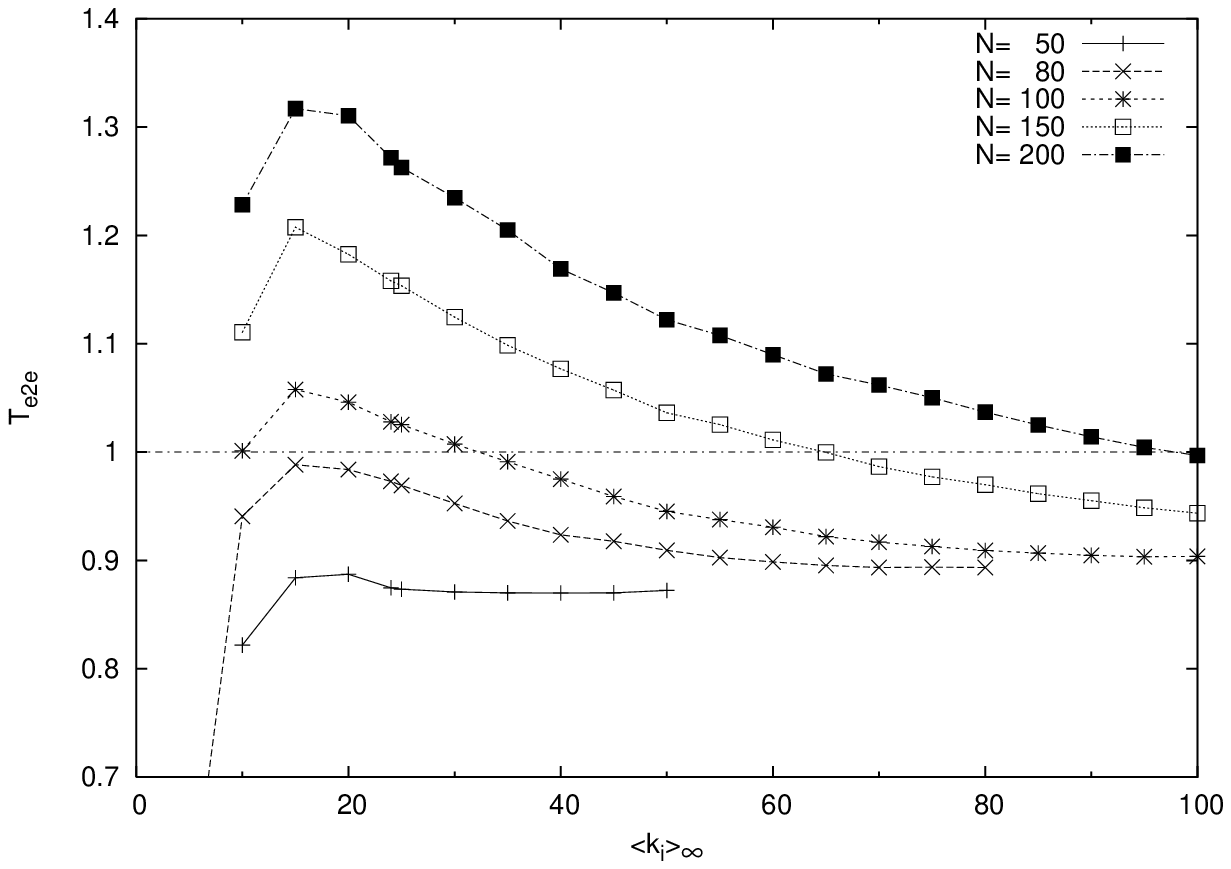,width=12cm}
\epsfig{file=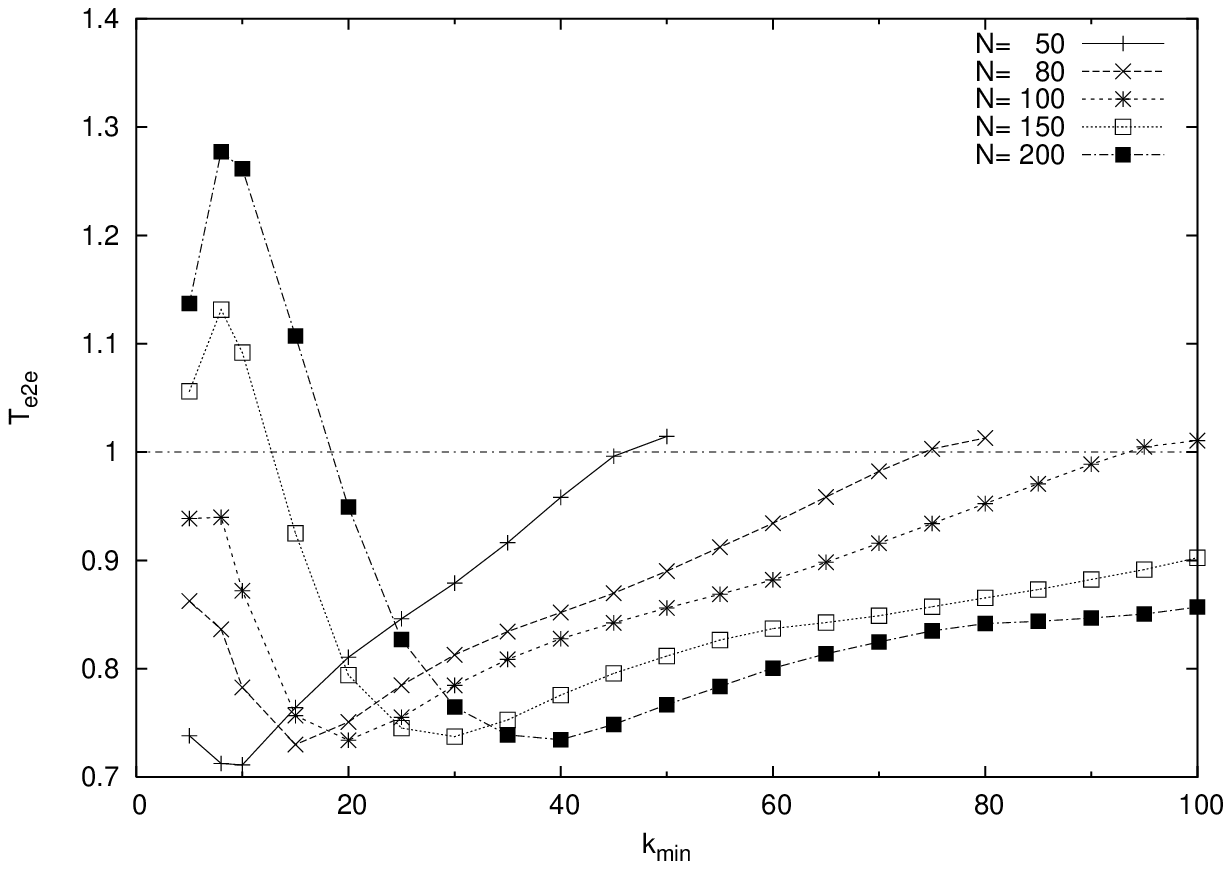,width=12cm}
\caption{
End-to-end throughput, obtained from generic data-traffic simulations, 
for models I (top) and II (bottom) for various values of $N$
as a function of $\langle k_i \rangle_\infty$ and $k_{min}$, respectively.
}
\label{fig:throughput_vs_degree}
\end{centering}
\end{figure}

However, this observational correlation conjecture turns out not to be
true. Fig.\ \ref{fig:throughput_vs_degree}b shows the network-model II 
end-to-end throughput for fixed network size $N$ as a function of the 
minimum node degree $k_{min}$. For $N\leq 100$, $T_{e2e}$ is below one 
for small $k_{min}$, then drops towards medium $k_{min}$ and finally 
increases to $T_{e2e}=1$ in the fully connected limit $k_{min}=N-1$. A 
similar conclusion is drawn from Fig.\ \ref{fig:throughput_vs_degree}a 
for network model I. Note there, that at 
$\langle k_i \rangle_\infty = N-1$ the end-to-end throughput has not 
converged to the fully connected $T_{e2e}=1$. This is a finite-size 
effect, because nodes close to the border of the chosen square area 
$L\times L$ are missing nodes at the other side to connect to. Their 
node degree is smaller than the asymptotic $N-1$, implying a network 
structure which is not fully connected. Only once all nodes come with 
a sufficiently larger transmission power, implying
$\langle k_i \rangle_\infty > N-1$, the end-to-end throughput converges
back to its fully connected limit.

Upon focusing on larger sized networks, the results exemplified in 
Figs.\ \ref{fig:Te2e}b and \ref{fig:throughput_vs_degree}a+b reveal
that sparsely connected networks then come with an end-to-end 
throughput, which is larger than the fully connected $T_{e2e}=1$. The
critical network size, which separates $T_{e2e}<1$ for $N<N_{crit}$
from $T_{e2e}>1$ for $N>N_{crit}$, is of the order of 
$N_{crit} \approx 100$. See also Table \ref{tab:Te2e}. In view of
several envisaged technological applications of small-sized wireless
multihop ad hoc networks, this is an important statement.

\begin{table}
\begin{tabular}{l|rrr|rrr}
$T_{e2e}(N)=a N^\gamma$   & \multicolumn{3}{c|}{model $\mathrm{I}_{24}$} 
                       & \multicolumn{3}{c}{model $\mathrm{II}_{8}$} \\
$T_{e2e}(N_{crit})\equiv 1$  & $N_{crit}$ & $a$   & $\gamma$ 
                       & $N_{crit}$ & $a$   & $\gamma$ \\ \hline
data traffic           &         89 & 0.167 & 0.383    
                       &        115 & 0.368 & 0.242 \\
Eq.\ (\ref{eq:tput1})  &         43 & 0.135 & 0.492    
                       &         39 & 0.131 & 0.535 \\
Eq.\ (\ref{eq:tput2b}) &        143 & 0.135 & 0.403    
                       &         97 & 0.398 & 0.233 \\
Eq.\ (\ref{eq:tput3}), $\Delta \tau_1$ 
                       &         63 & 0.179 & 0.385    
                       &         59 & 0.466 & 0.225 \\
Eq.\ (\ref{eq:tput3}), $\Delta \tau_1 + \Delta \tau_2$ 
                       &        178 & 0.145 & 0.368    
                       &        128 & 0.337 & 0.232 \\
\end{tabular}
\vskip1ex
\caption{
End-to-end throughput $T_{e2e}=a N^\gamma$ and critical size
$T_{e2e}(N_{crit})\equiv 1$ for network models I with
$\langle k_i \rangle = 24$ and II with $k_{min}=8$ 
obtained from generic data traffic simulations and the estimates
(\ref{eq:tput1}), (\ref{eq:tput2b}) and (\ref{eq:tput3}). For the last
estimate two variations have been employed: the first one uses 
(\ref{eq:tsend12}) without the $\Delta \tau_2$ term and the second one
uses the full expression (\ref{eq:tsend12}). The parameters $a$ and
$\gamma$ have been obtained from best fits in the regime 
$200\leq N\leq 2000$.
}
\label{tab:Te2e}
\end{table}

For network sizes larger than $N_{crit}$ the end-to-end throughput is 
larger than one and increases the larger $N$ becomes. In this regime, 
scalability statements are of utmost importance. For network models I 
and II such statements can be given; see \figref{fig:Te2e}b. Within 
$200\leq N\leq 2000$ the scaling expression $T_{e2e}=a N^\gamma$ is 
found to give a precise agreement with the generic-data-traffic 
results. The fitted parameters are listed in the first row of Table 
\ref{tab:Te2e}. The scaling exponents are found to be $\gamma=0.38$ 
for network model I with $\langle k_i \rangle_\infty = 24$ and 
$\gamma=0.24$ for network model II with $k_{min}=8$. This outcome 
clearly shows, that network structure has a prominent influence on the 
network performance with respect to data traffic. It is the goal of the 
next Section to present semi-analytic estimates, which are compatible 
with the e2e-throughput scalability observed from the data-traffic 
simulations.

\section{Modeling of end-to-end throughput}
\label{sec:meanfield}

\subsection{Throughput I: back-on-the-envelope estimate}
\label{subsec:throughputI}

The end-to-end throughput describes the maximum number of end-to-end
communications, which can be completed per time step without network
overloading. This maximum number can be roughly estimated as the maximum
number of one-hop transmissions taking place per time step, divided by 
the average length $D$ of an end-to-end communication route:
\begin{equation}
\label{eq:tput1}
  T_{e2e} 
    \approx  \frac{1}{D} 
             \frac{N}{\left( 2 + \langle k_{i\leftrightarrow j}^{out} 
                      \rangle \right)}
             \; .
\end{equation}
On average, $2+\langle k_{i\leftrightarrow j}^{out} \rangle$ nodes
are involved in a one-hop transmission, the one-hop
sender and receiver as well as their $k_{i\leftrightarrow j}^{out}$
outgoing neighbors, which are blocked by medium access control. The 
network size $N$ divided by this number represents the maximum number 
of simultaneous one-hop transmissions taking place per time step. 
--
The expression (\ref{eq:tput1}) is consistent with the finding 
$T_{e2e}=1$ for fully connected networks. Since every node has a 
direct link to every other node, we have $D=1$ and 
$k_{i\leftrightarrow j}^{out}=N-2$.

\begin{figure}
\begin{centering}
\noindent
\begin{tabular}{cc}
\epsfig{file=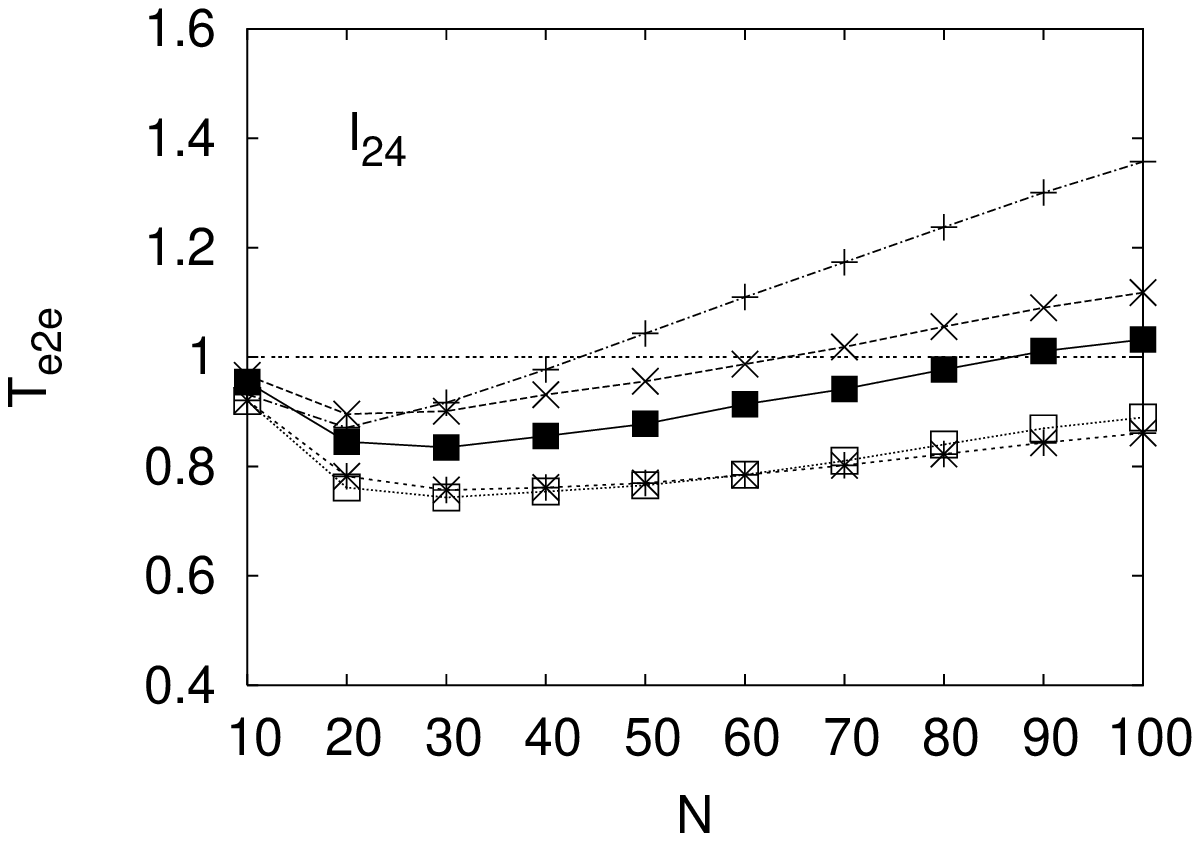,width=6.5cm} &
\epsfig{file=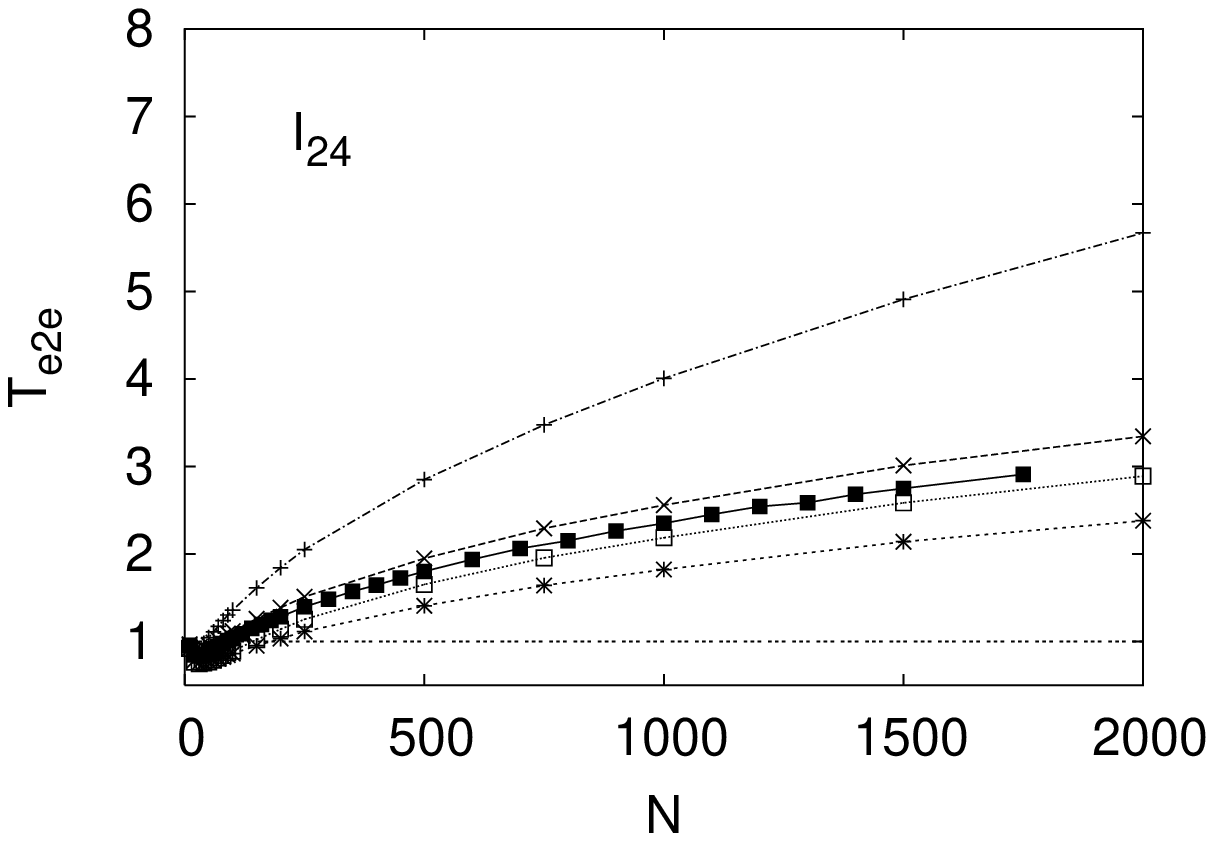,width=6.5cm} \\
\epsfig{file=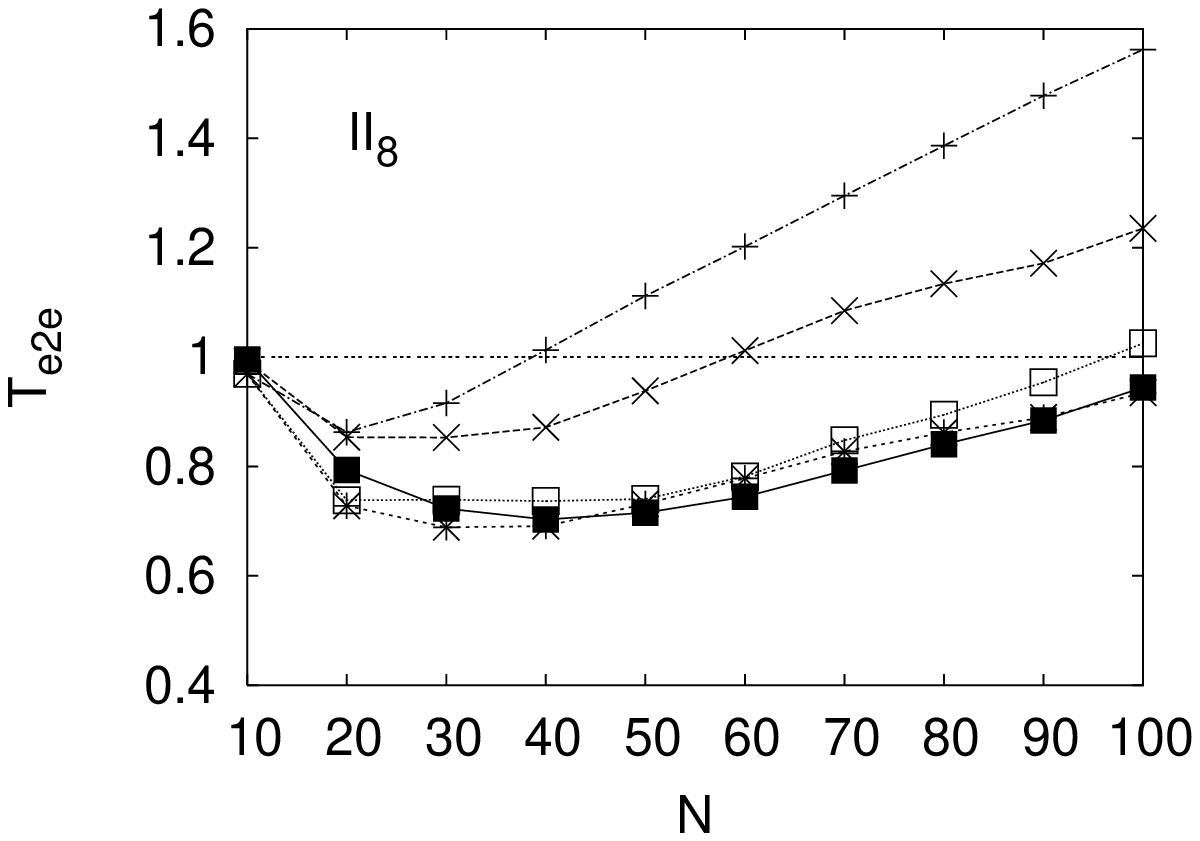,width=6.5cm} &
\epsfig{file=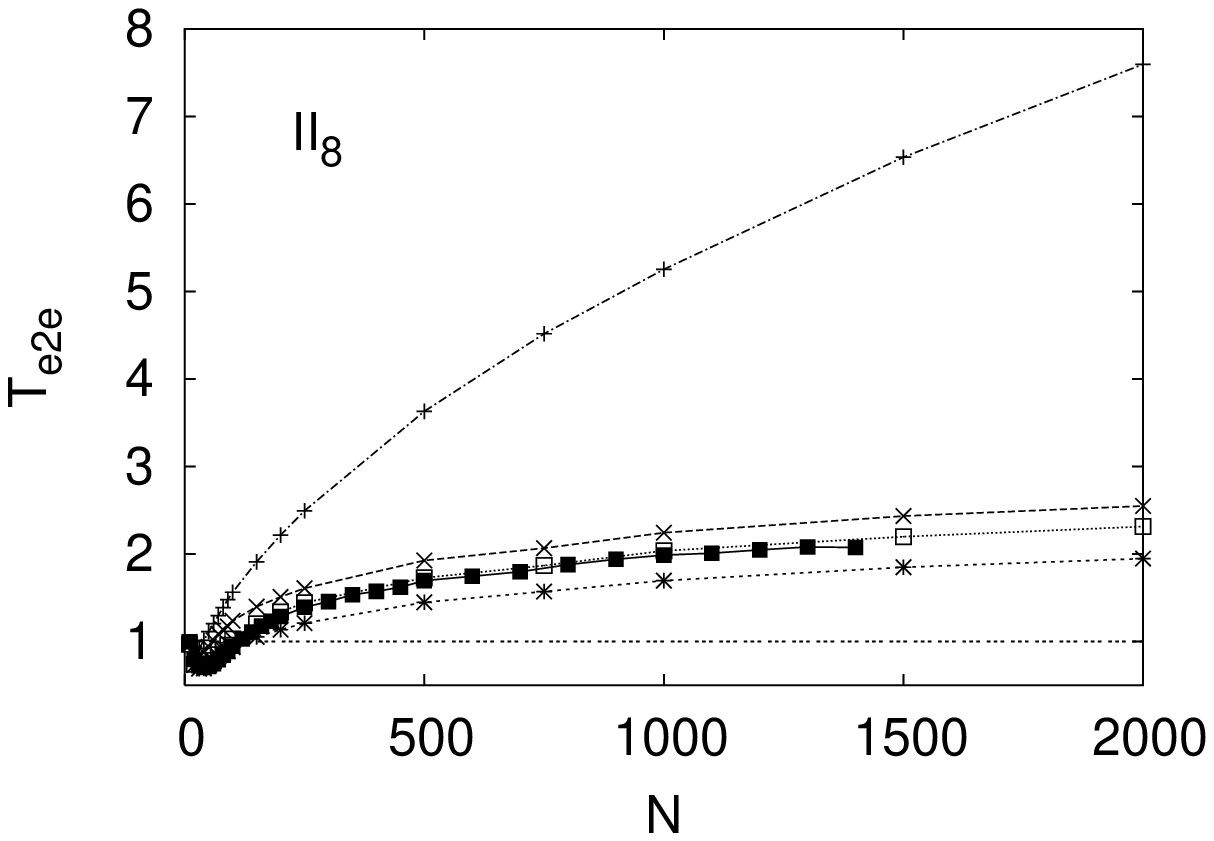,width=6.5cm} \\
\epsfig{file=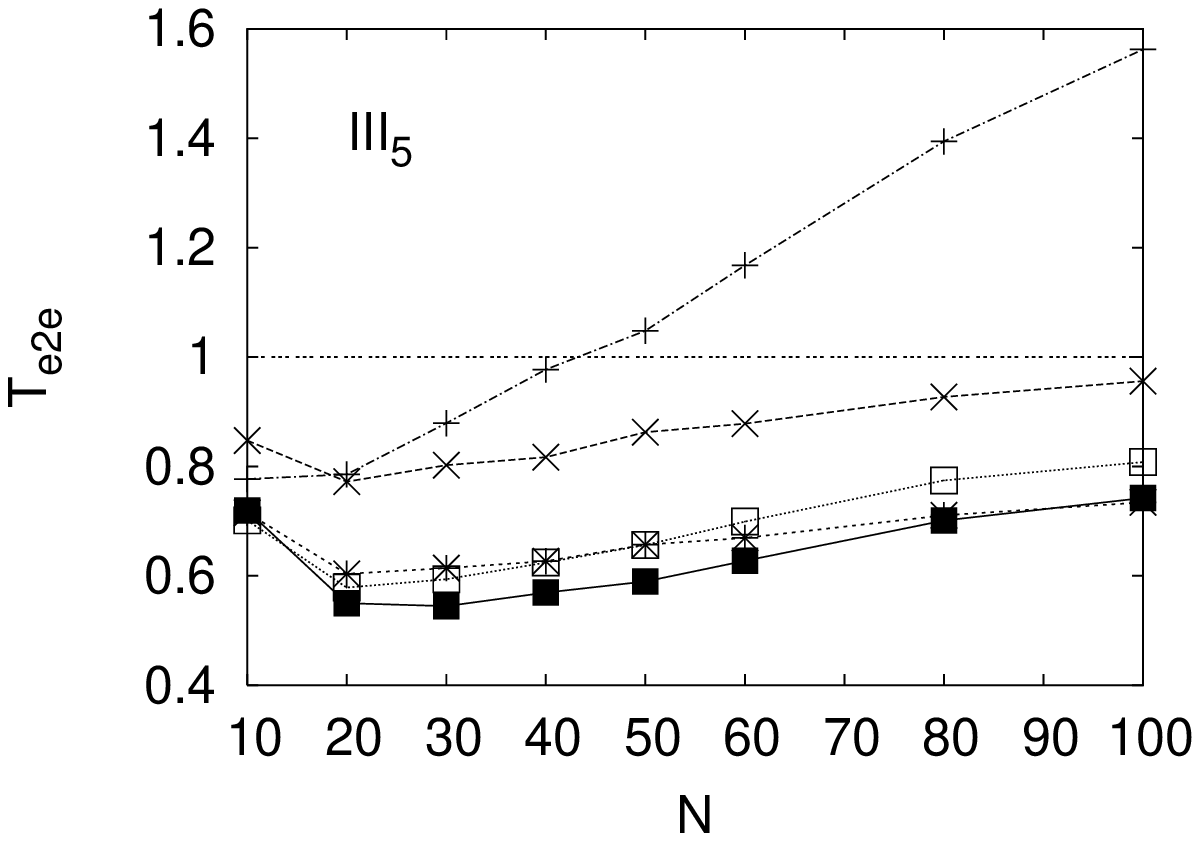,width=6.5cm} &
\epsfig{file=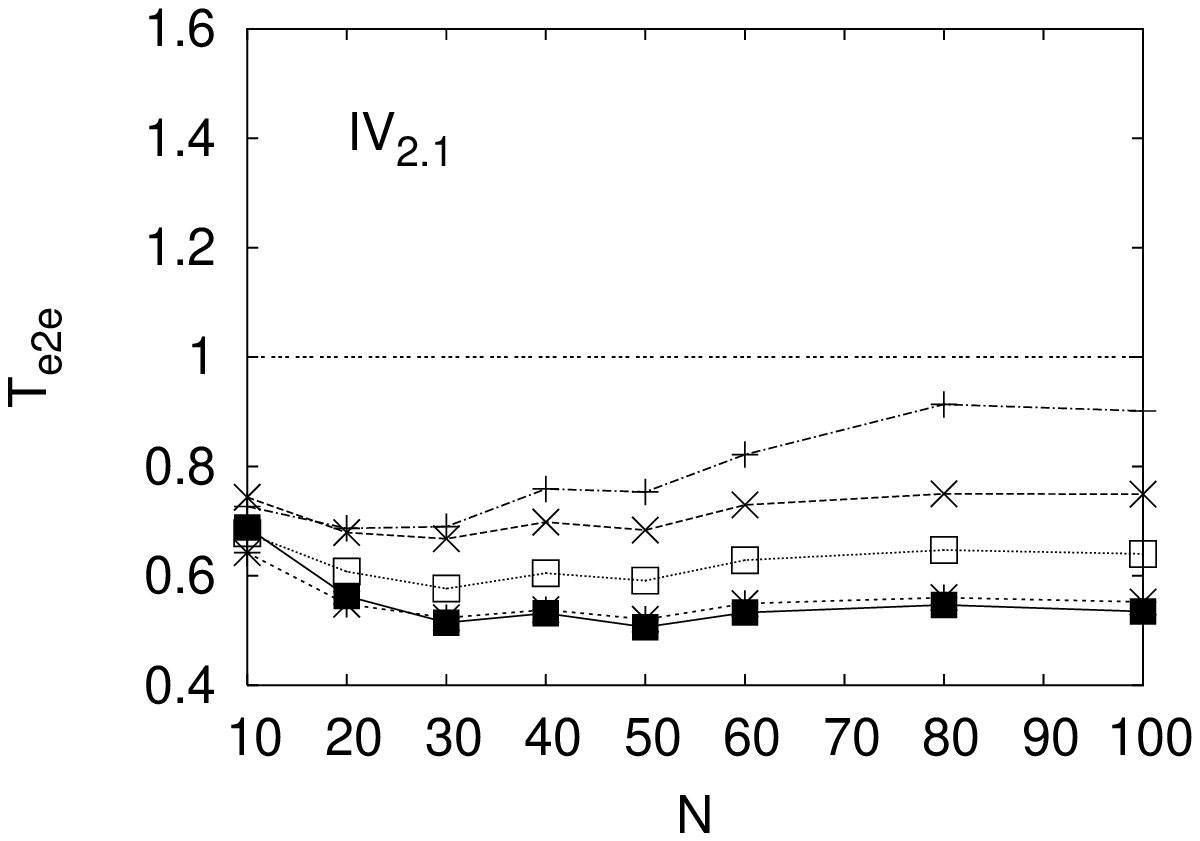,width=6.5cm} \\
\epsfig{file=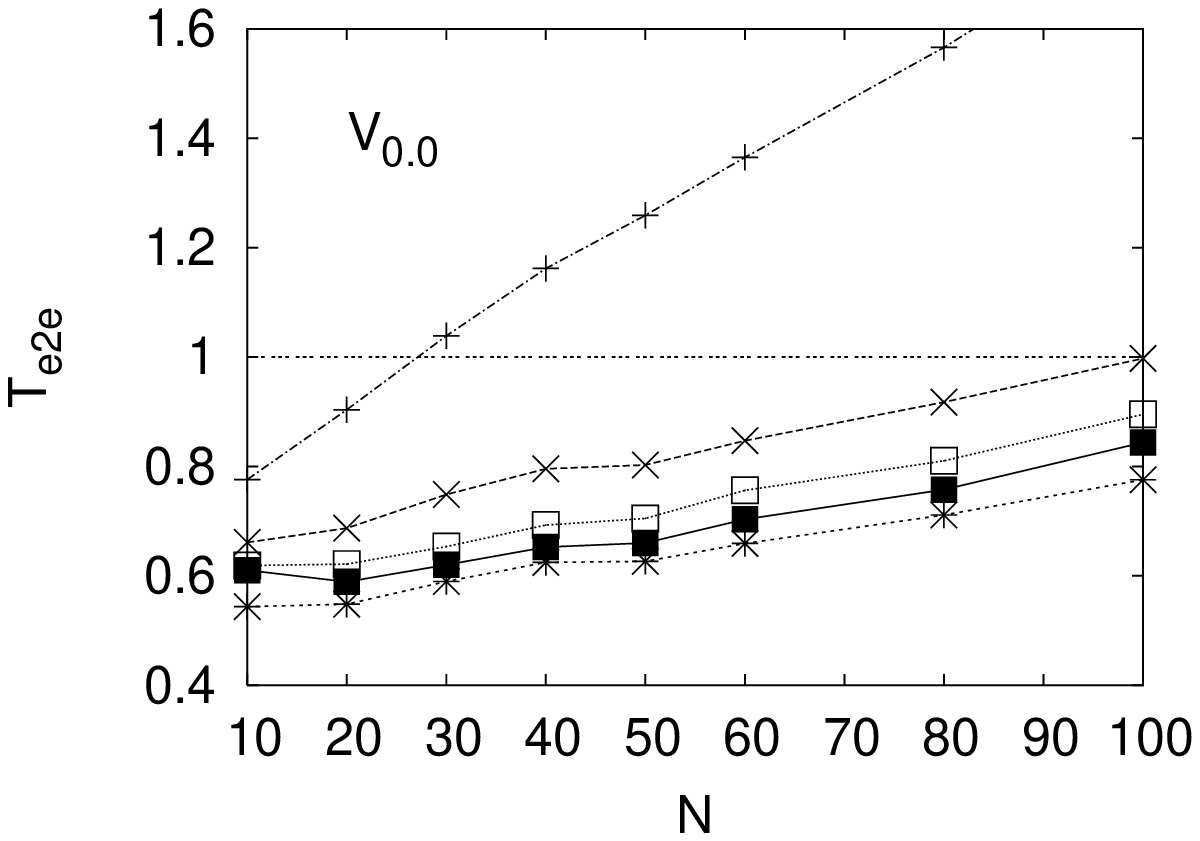,width=6.5cm} &
\epsfig{file=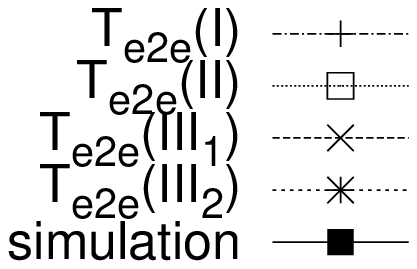,width=4cm} \\
\end{tabular}
\end{centering}
\caption{
Comparison of the $N$-dependent end-to-end throughput obtained from the 
generic data traffic simulations (filled squares) and the estimates
(\ref{eq:tput1}) (vertical crosses), 
(\ref{eq:tput2b}) (open squares),
(\ref{eq:tput3}) without $\Delta\tau_2$ in (\ref{eq:lineqs}) 
(rotated crosses),  and
(\ref{eq:tput3}) with $\Delta\tau_2$ in (\ref{eq:lineqs}) (stars).
The used network models are 
I (top row) with $\langle k_i \rangle = 24$,
II (second row) with $k_{min}=8$,\
III (left part of third row) with $k_{target}=5$,
IV (right part of third row) with $\gamma=2.1$, and
V (last row) with $\lambda=0$.
}
\label{fig:Te2e_gt}
\end{figure}

For the various network models the expression (\ref{eq:tput1}) can be
determined right away. \figref{fig:Te2e_gt} compares the sample average
with the results from the generic data traffic simulations and reveals 
that for all network models I--V the former noticeably overestimates the 
latter. This is not the only discrepancy. A quick look into the second
row of Table \ref{tab:Te2e} reveals that according to (\ref{eq:tput1})
the end-to-end throughput of network models I and II would scale with
an exponent very close to $1/2$. This outcome, being a consequence of 
the more or less universal $D\sim\sqrt{N}$ behavior, has already been 
proposed earlier \cite{GUP00}, but is in conflict with the findings 
from the generic data traffic simulations. The inherent difficulty with
the at first plausible estimate (\ref{eq:tput1}) is that it is built 
with average network properties. Contrary to the generic data traffic 
it does not account for the most-critical-node effect, which limits
the end-to-end throughput.

\subsection{Throughput II: cumulative node inbetweeness}
\label{subsec:throughputII}

Analogous to the most-critical-node procedure 
(\ref{eq:vierb1})-(\ref{eq:vierb3}) for the determination of the 
end-to-end throughput (\ref{eq:vierb4}) of the generic data traffic,
a more sophisticated estimate for $T_{e2e}$ is now given. The mean
in-flux of data packets into node $i$,
\begin{equation}
\label{eq:muinB}
  \mu^{in}_i
    =  \mu N \frac{B_i}{N(N-1)}
       \; ,
\end{equation}
is given by the fraction $B_i/N(N-1)$ of the $\mu N$ newly created data 
packets per time step, that will be routed via node $i$ at later time 
steps in order to reach their final destination. Within the numerical
uncertainty of the generic data traffic simulations we have tested that
the two flux rates (\ref{eq:vierb1}) and (\ref{eq:muinB}) are indeed
identical. A realistic description of the out-flux (\ref{eq:vierb2}) 
is subject to modeling. A particular simple estimate of the average 
sending time uses the cumulative node inbetweeness (\ref{eq:zweib7}):
\begin{equation}
\label{eq:tput2a}
  \tau_i 
    =  \frac{B_i^{cum}}{B_i}
    =  1 + \frac{1}{B_i} \sum_{j\in{\mathcal N}_i^{in}} B_j
       \; .
\end{equation}
As node $i$ competes with its ingoing neighbors for medium access, it
may have to wait a certain amount of time before sending its packet
along a one-hop connection. The ansatz (\ref{eq:tput2a}) weights each
of the competing nodes with its relative frequent use for forwarding
end-to-end communications. It represents an interpolation between 
$\tau_i \rightarrow 1$ in the low-load limit $\mu\rightarrow 0$ and 
$\tau_i \rightarrow \mathcal{O}(k_i^{in})$ holding in the supercritical 
regime $\mu\gg\mu_{crit}$. As such, it can be hoped that the ansatz 
(\ref{eq:tput2a}) catches at least part of the truth when it comes to 
the proper behavior of the sending time around the critical network 
load $\mu_{crit}$. Using (\ref{eq:muinB}) and (\ref{eq:tput2a}), the
end-to-end throughput (\ref{eq:vierb4}) now becomes
\begin{equation}
\label{eq:tput2b}
  T_{e2e} 
    \approx  \frac{N(N-1)}{\langle \sup_i B_i^{cum} \rangle}
             \; .
\end{equation}
Like (\ref{eq:tput1}), the expression (\ref{eq:tput2b}) is consistent 
with the finding $T_{e2e}=1$ for fully connected networks. In this case 
every node has the same node inbetweeness $B_i=N-1$, which results in 
$B_i^{cum}=N(N-1)$ for the cumulative node inbetweeness.

As can be read off from \figref{fig:Te2e_gt} the estimate 
(\ref{eq:tput2b}) is closer to the generic-data-traffic e2e-throughput
than the previous estimate (\ref{eq:tput1}). For some of the 
$N\leq 100$ network models, like II, III and V, it even almost matches 
its data-traffic counterpart. A similar statement can be given for 
network models I and II in the $100\leq N\leq 2000$ regime; compare 
also the first and third rows of Table \ref{tab:Te2e}. According to the 
expression (\ref{eq:tput2b}) the scaling exponents of 
$T_{e2e} \sim N^\gamma$ are directly determined from the scaling 
behavior of $\langle \sup_i B_i^{cum} \rangle$, which has already been 
summarized in Table \ref{tab:fit_2000}, and are almost identical to the 
corresponding data-traffic results. In retrospect, this represents 
a justification for the estimate (\ref{eq:tput2b}).

\subsection{Throughput III: sending-times estimate}
\label{subsec:sendtimes}

The modeling (\ref{eq:tput2a}) of the average sending time does not 
account for its proper dependence on the packet creation rate $\mu$. 
This shortcoming will now be removed. 

Let us assume that node $i$ has at least one packet to forward. If all 
nodes in its neighborhood have no packets at this instance, then there
is no competition for medium access and node $i$ can send its packet 
right away. In this case $\tau_i = 1$. Competition sets in once some 
of the one-hop neighbors also have packets to forward. In addition, also 
some of the two-hop neighbors enter the competition, once they have 
packets to transmit to one-hop neighbors of $i$. This brings us to the
structural ansatz:
\begin{equation}
\label{eq:tsend12}
  \tau_i 
    =  1 + \Delta\tau_1 + \Delta\tau_2
           \; .
\end{equation}
 
For simplification, we neglect spatio-temporal correlations of the
network traffic. The independent-node picture then leads to
\begin{equation}
\label{eq:tsend1}
  \Delta\tau_1
    =  \sum_{j_{1}\in\mathcal{N}_i^{in}} p_{j_1}(n_{j_{1}} \ge 1)
\end{equation}
for the competition term from the one-hop neighbors. It sums over the 
probabilities of ingoing one-hop neighbors to have at least one packet
($n_{j_{1}} \ge 1$) to forward. On the same footing, the competition 
term from the two-hop neighbors results to be 
\begin{equation}
\label{eq:tsend2}
  \Delta\tau_2
    =  \sum_{j_{2}\in\mathcal{N}(\mathcal{N}_i^{in})
             \backslash\mathcal{N}_i^{in}}
       p_{j_2}(n_{j_{2}} \ge 1) 
       \sum_{j_{1}\in\mathcal{N}_i^{in}}
       \frac{B_{j_{2}\leftrightarrow j_{1}}}{2B_{j_{2}}}
       \; .
\end{equation}
Here it is not sufficient for a two-hop neighbor $j_2$ to have at least 
one packet, the packet also needs to be forwarded to a one-hop neighbor
$j_1$ in order to compete with $i$ for medium access. The ratio 
$B_{j_{2}\leftrightarrow j_{1}} / 2B_{j_{2}}$ represents this fraction.

Note, that with (\ref{eq:tsend2}) the competition strength of the 
two-hop neighbors is overestimated. Active three hop-neighbors might 
have already blocked some of the two-hop neighbors, so that the sum 
over $j_2$ should not extend to the complete two-hop neighborhood of 
node $i$. To a minor fraction this overestimation is also true for the 
expression (\ref{eq:tsend1}), since a competing one-hop neighbor might 
already be blocked by an earlier assigned transmission between two 
two-hop neighbors. Consequently, the expressions (\ref{eq:tsend1}) and 
(\ref{eq:tsend2}) should be corrected by the probability that the one- 
and two-hop neighbor is not blocked, respectively. For the moment this 
is too much of a complication. From now on we will operationally 
proceed with two versions of (\ref{eq:tsend12}), one excluding and the 
other including the $\Delta\tau_2$ contribution (\ref{eq:tsend2}).

We are left to model the probability distributions $p(n_i)$ of the 
queue lengths in the independent-node picture. It is sufficient to 
characterize the single-node data traffic statistics by the average 
in- and out-flux. In the stationary limit, this immeadiately leads to 
the rate equation
\begin{equation}
\label{eq:rateeq}
  \left( \mu^{in}_i + \mu^{out}_i \right) p_i(n_i) 
    =  \mu^{in}_i p_i(n_i-1) + \mu^{out}_i p_i(n_i+1)
       \; .
\end{equation}
Its solution leads to 
\begin{equation}
\label{eq:pdfneq0}
  p_i(n_i\geq 1) 
    =  \frac{\mu^{in}_i}{\mu^{out}_i}
       \; .
\end{equation}
Making again use of (\ref{eq:vierb2}) and (\ref{eq:muinB}), the 
proposed relation (\ref{eq:tsend12}) for the average sending time then 
transforms into an inhomogeneous set of $N$ coupled linear equations:
\begin{eqnarray}
\label{eq:lineqs}
  \tau_i 
    &=&  1 + 
         \sum_{j_{1}\in\mathcal{N}_i^{in}} 
         \frac{\mu B_{j_1}}{(N-1)} \tau_{j_1} 
         \nonumber \\
    & &  + \sum_{j_2\in\mathcal{N}(\mathcal{N}_i^{in})
                 \backslash\mathcal{N}_i^{in}}
         \frac{\mu B_{j_2}}{(N-1)} \tau_{j_2} 
         \sum_{j_{1}\in\mathcal{N}_i^{in}}
         \frac{B_{j_2\leftrightarrow j_1}}{2B_{j_2}}
         \; .
\end{eqnarray}
The solution for the node-dependent average sending times will 
explicitly depend on the packet creation rate $\mu$ in a nonlinear way.

Let us stress this last point by picking a fully connected network as
an example. Since every node is already a one-hop neighbor to each
other, the last term on the righthand side of (\ref{eq:lineqs}) has to
be dropped. Furthermore, $B_i=N-1$ for every node and by symmetry all
$\tau_i=\tau$ are identical. This transforms (\ref{eq:lineqs}) into 
$\tau =(1-\mu(N-1))^{-1}$, which is a nonlinear function of the packet 
creation rate. Let us continue for a moment, use the last result for 
the out-flux (\ref{eq:vierb2}) and equate it to the in-flux 
(\ref{eq:muinB}), then as expected the critical network load results 
in $\mu_{crit}=1/N$. Upon inserting this into the expression for
the sending time, we arrive at $\tau(\mu_{crit})=N$, which
again reflects our intuition that at the critical network load every
node has to compete with every other node for medium access.

Of course for other, more complicated network structures a further
analytical insight into (\ref{eq:lineqs}) is hard to give. However,
once a specific network realization of size $N$ is constructed, the 
numerically accessible node and link inbetweeness variables serve as 
input and the coupled linear equations can be solved numerically as 
a function of $\mu$. 

The solution of (\ref{eq:lineqs}) fixes the out-flux (\ref{eq:vierb2}). 
Upon setting this equal to the in-flux (\ref{eq:vierb1}), the 
node-specific critical packet creation rate is determined to 
$\mu^{crit}_i = (N-1) / B_i \tau_i(\mu^{crit}_i)$. Via 
(\ref{eq:vierb3}) and (\ref{eq:vierb4}) this finally leads to the 
expression
\begin{equation}
\label{eq:tput3}
  T_{e2e} 
    =  \frac{N(N-1)}{\left\langle \sup_i B_i \tau_i(\mu^{crit}_i)
                     \right\rangle}
\end{equation}
for the end-to-end throughput. For consistency, we again check on fully 
connected networks with $B_i=N-1$ and $\tau_i(\mu^{crit}_i) = N$, which 
leads to $T_{e2e}=1$.

Depending on whether the two-hop contribution $\Delta\tau_2$ is ex- or
included in (\ref{eq:lineqs}), two further estimates for the end-to-end
throughput can be given for network models I--V. They are both illustrated
in \figref{fig:Te2e_gt}. The $\Delta\tau_2$-excluding estimate yields an
end-to-end throughput a little larger than the result obtained from the
generic data traffic simulations. Except for network model I, the 
$\Delta\tau_2$-including estimate produces a remarkable agreement with the
generic data-traffic observations for small network sizes. For larger
network sizes of models I and II the two estimates with and without
$\Delta\tau_2$ narrowly sandwich the generic data-traffic outcome.
Moreover, the fourth and fifth row of Table \ref{tab:Te2e} confirm that 
the corresponding scaling exponents for $T_{e2e}\sim N^\gamma$ are in 
close agreement with those found from the generic data traffic simulations. 
All of this clearly demonstrates that the deviations from the previously 
proposed $\gamma=1/2$ \cite{GUP00} are a consequence of the 
most-critical-node effect and that these deviations do show a strong 
dependence on the underlying network structure.

\section{Conclusion and Outlook}
\label{sec:conclusion}

Wireless ad hoc communication networks represent an example for a
complex, networked technological system. Their key operational 
controls, routing and medium access control, come with opposite demands 
and leave the network in a state of frustration. This opens the door 
for the Statistical Physics of complex networks to ask, what are 
efficient network structures in order to find a good compromise. 
Various network models, all of them constrained by spatial geometry, 
the wireless propagation medium and some selected design principles, 
have been proposed to explore the huge network structure state space 
and to test the impact of network structure on network performance. As 
a measure the end-to-end throughput has been chosen. It represents the 
network's capacity, sometimes also called bandwidth, to deliver a 
certain amount of end-to-end communications per time without network 
overloading. A generic data traffic simulation as well as several 
semianalytic estimates of increasing sophistication demonstrate that 
(i) below a critical network size a fully connected structure always
comes with the largest and then size-independent throughput,
(ii) well above the critical network size the throughput scales as 
$T_{e2e} \sim N^\gamma$ with network size $N$,
(iii) contrary to current mean-field belief \cite{GUP00} 
$\gamma \neq 0.5$ due to the most-critical node effect, and
(iv) the scaling exponent $\gamma$ depends on the specific structure
of the picked network model.
Another, although smaller discovery along the way has been that 
different network structure models lead to quite different average 
transmission power values, but once the latter is weighted with the 
amount of data traffic passing over a node, the differences in power 
consumption between the network structure models decline. 

The generic network approach to wireless ad hoc communication, which 
has been advocated in this Paper, gives guidance to several yet-to-come 
generalizations. Straightforward issues like the spatial heterogeneity 
of static and mobile point patterns, the spatio-temporal heterogeneity 
of the wireless propagation medium, interference, synchronization and 
different user behavior for the modeling of data traffic need to be 
addressed. Additional network structure models, directly resulting from 
the global optimization of design principles, like the maximization of 
end-to-end throughput itself, need to be developed and will then serve 
as benchmarks for an even bigger challenge, the construction of such 
optimized network structures via decentralized topology control rules. 
Functional aspects like routing coupled to congestion control and new
technology-driven aspects like geometric routing \cite{KUH03} are then 
also becoming important. This underlines that it is still a longer way 
to reach a truly selforganizational control of wireless ad hoc networks 
and other networked technological systems. Upon further pursuing this 
road, it is a good idea to learn more about selforganization also from 
other networked complex systems out of physics, biology and sociology.

\section{Appendix: graph-theoretical variables}
\label{sec:appendix}

\paragraph*{Node degree}
The node degree $k_i=\sum_{j=1}^N a_{ij}a_{ji}$ counts the number of 
bidirected neighbors of node $i$. The $N{\times}N$ adjacency matrix
$a_{ij}=1$ (link $i{\rightarrow}j$ exists) or $0$ (link 
$i{\rightarrow}j$ does not exist, or $i=j$) most naturally embraces 
the existence and nonexistence of links between nodes. The node degree 
is smaller or equal to the outgoing node degree 
$k_i^{out}=\sum_{j=1}^N a_{ij}$ or the ingoing node degree
$k_i^{in}=\sum_{j=1}^N a_{ji}$. The sets of bidirected, outgoing and 
ingoing neighbors of node $i$ are denoted as ${\mathcal N}_i$, 
${\mathcal N}_i^{out}$ and ${\mathcal N}_i^{in}$, respectively.

\paragraph*{Link degree}
Using the definition $\mathcal{N}_{i{\leftrightarrow}j} =
\left( \mathcal{N}_i \cup \mathcal{N}_j \right) \setminus \{i,j\}$
for all bidirected links $i{\leftrightarrow}j$, the link degree
$k_{i{\leftrightarrow}j} = |{\mathcal N}_{i{\leftrightarrow}j}|$ is 
similar to the node degree $k_i$. It counts the number of bidirected 
neighbors, which are attached either to node $i$ or to node $j$ or to 
both.  Two straightforward and selfexplaining generalizations are 
$k_{i{\leftrightarrow}j}^{out}$ and $k_{i{\leftrightarrow}j}^{in}$. 
They are of direct relevance for wireless multihop ad hoc networks.  
$k_{i{\leftrightarrow}j}^{out}$ counts all outgoing neighboring nodes 
of the active one-hop link $i{\leftrightarrow}j$, which are silenced 
by medium access control, and $k_{i{\leftrightarrow}j}^{in}$ represents 
the maximum number of possible data packet senders, which directly 
compete with nodes $i$ and $j$ for medium access. 

\paragraph*{Cluster coefficient (node-based)}
The cluster coefficient  
$C_{i} = 
(2/[k_i(k_i-1)]) \sum_{j_1>j_2\in{\mathcal N}_i} a_{j_1j_2} a_{j_2j_1}$
is another popular observable \cite{WAT98}. It counts the number of 
existing bidirected links with respect to all possible bidirected links 
among the nodes belonging to the bidirected neighborhood 
${\mathcal N}_i$ of node $i$. One-directed variants of the cluster 
coefficient are not considered. 

\paragraph*{Cluster coefficient (link-based)}
Analogous to the distinction between node and link degree, the 
cluster coefficient can not only be defined in the node-based manner, 
but also in the link-based manner 
$C_{i{\leftrightarrow}j} =
(2/[k_{i{\leftrightarrow}j}(k_{i{\leftrightarrow}j}-1)])
\sum_{m_1>m_2\in{\mathcal N}_{i{\leftrightarrow}j}}
a_{m_1m_2} a_{m_2m_1}$.
It counts the number of existing bidirected links with respect to all 
possible bidirected links among the nodes belonging to the bidirected 
neighborhood ${\mathcal N}_{i{\leftrightarrow}j}$ of link 
$i{\leftrightarrow}j$. 

\paragraph*{Diameter}
The shortest-path distance $d_{ij}$ counts the number of bidirected 
hops along the shortest (in units of hops) path between nodes $i$ and 
$j$. The diameter $D = \langle d_{ij} \rangle$ is given by double 
averaging, in the first step a mean over all node pairs of one network 
realization and in the second step a mean over a sample of network 
realizations.

\paragraph*{Node Inbetweeness}
For simplicity, we assume table-based shortest-path routing using 
bidirected links for all possible end-to-end communications within the 
network. In general different nodes will be used with different 
frequency for the relaying of packets. A measure of this frequency is 
given by the node inbetweeness $B_i$, which counts the number of 
shortest paths out of the $N(N-1)$ different shortest end-to-end 
communication routes going via node $i$. To be more precise, all nodes 
that are subsequently transmitting along such a shortest path, add a 
one to their respective node-inbetweeness counter; this includes the 
initially transmitting node, but not the final recipient. If between 
two ends more than one, say $n$ shortest paths exist, then each path is 
weighted with $1/n$ and nodes along such a path add this weight to 
their node-inbetweeness counter. This definition of node inbetweeness 
is similar to the well known betweenness centrality \cite{NEW01}. 
Contrary to the former, the latter excludes contributions from the 
initial and final node. For the calculation of the node inbetweeness we 
use an all-pairs shortest path algorithm similar to that described in 
Ref.\ \cite{NEW01}. The sum rule
\begin{equation}
\label{eq:zweib4}
  N \langle B_i \rangle
    =  N(N-1) D
\end{equation}
relates the average node inbetweeness to the network diameter.

\paragraph*{Link Inbetweeness}
Similar to the definition of node inbetweeness the link inbetweeness
is a measure for the importance of a link. The link inbetweeness
$B_{i{\leftrightarrow}j}$ counts the number of shortest paths that
contain the link $i{\leftrightarrow}j$. If there exist $n$ shortest
paths between two nodes, we proceed as described above for the node
inbetweeness, where each of these paths gets the weight $1/n$. The link
inbetweeness is connected with the node inbetweeness by the flux rule
\begin{equation}
\label{eq:zweib5}
  B_{i}  =  \frac{1}{2} \sum_{j\in{\mathcal N}_i} 
            B_{i{\leftrightarrow}j}
            \; ,
\end{equation}
which allows to express the mean link inbetweeness 
\begin{equation}
\label{eq:zweib6}
  \left\langle B_{i{\leftrightarrow}j} \right\rangle
    =  \frac{ 2 (N-1) D }{ \langle k_i \rangle }
\end{equation}
in terms of the network diameter and the mean node degree.

\paragraph*{Cumulative Node Inbetweeness}
For our discussion in Sect.\ \ref{sec:meanfield} on the 
end-to-end throughput the quantity
\begin{equation}
\label{eq:zweib7}
  B_i^{cum} 
     =  B_i + \sum_{j\in{\mathcal N}_i^{in}} B_j
        \; ,
\end{equation}
is of relevance. We call it the cumulative node inbetweeness, since it 
adds the node inbetweennesses of all ingoing neighbors to the node 
inbetweeness of node $i$. The second term on the right-hand side of 
(\ref{eq:zweib7}), when divided by $B_i$, can be seen as an 
inbetweeness-weighted ingoing node degree. By reordering, the mean can 
be expressed as 
$\langle B_i^{cum} \rangle = \langle B_i (1+k_i^{out}) \rangle$.

\ack
W.\ K.\ acknowledges support from the Ernst von Siemens-Scholarship.



\begin{thebibliography}{00}

\bibitem{ALB02}
  R.\ Albert and A.L.\ Barabasi,
  Rev.\ Mod.\ Phys.\ 74 (2002) 47.
\bibitem{DOR03}
  S.N.\ Dorogovtsev and J.F.F.\ Mendes,
  {\em Evolution of Networks -- From Biological Nets to the Internet 
  and WWW},
  Oxford University Press, Oxford (2003).
\bibitem{NEW03}
  M.E.J.\ Newman,
  SIAM Review 45 (2003) 167.
\bibitem{ALB03}
  R.\ Albert and H.\ Othmer,
  J.\ Theo.\ Bio.\ 223 (2003) 1.
\bibitem{COH02}
  R.\ Cohen, S.\ Havlin and D.\ ben-Avraham,
  Phys.\ Rev.\ Lett.\ 91 (2003) 247901.
\bibitem{FAL99}
  C.\ Faloutsos, P.\ Faloutsos and M.\ Faloutsos, 
  {\em On Power-Law Relationships of the Internet Topology},
  in Proc.\ ACM SIGCOMM (September 1999).
\bibitem{CHE02}
  Q.\ Chen, H.\ Chang, R.\ Govindan, S.\ Jasmin, S.\ Shenker 
  and W.\ Willinger,
  {\em The origin of power laws in Internet topologies revisited},
  in Proc.\ IEEE Infocom, New York, NY (June 2002).
\bibitem{CHA03}
  H.\ Chang, S.\ Jasmin and W.\ Willinger,
  {\em What Causal Forces Shape Internet Connectivity at the AS-level?},
  Technical Report CSE-475-03, EECS Dept., Univ.\ of Michigan (2003),
  http://citeseer.nj.nec.com/chang03what.html.
\bibitem{FUK99}
  H.\ Fuks and A.\ Lawniczak,
  Math.\ Comp.\ Sim.\ 51 (1999) 101. 
\bibitem{FUK01}
  H.\ Fuks, A.\ Lawniczak and S.\ Volkov,
  ACM Trans.\ Mod.\ Comp.\ Sim.\ 11 (2001) 233.
\bibitem{SOL01}
  R.\ Sole and S.\ Valverde,
  Physica A 289 (2001) 595.
\bibitem{VAL02}
  S.\ Valverde and R.\ Sole,
  Physica A 312 (2002) 636.
\bibitem{PAR00}
  K.\ Park and W.\ Willinger (eds),
  {\em Self-similar network traffic and performance evaluation},
  John Wiley \& Sons, New York (2000).
\bibitem{TOR03}
  Z.\ Toroczkai, G.\ Korniss, M.A.\ Novotny and H.\ Guclu,
  {\em Virtual time horizon control via communication network design},
  arXiv:cond-mat/0304617.
\bibitem{MANET}
  Mobile Ad Hoc Networks (manet) Working Group, \\
  http://www.ietf.org/html.charters/manet-charter.html.
\bibitem{NISTB}
  Wireless Ad Hoc Networks Bibliography, \\
  http://w3.antd.nist.gov/wctg/manet/manet{\_}bibliog.html.
\bibitem{MOB02}
  MobiHoc 2002,
  Proc.\ of 3rd ACM Int.\ Symp.\ on Mobile Ad Hoc Networking and
  Computing, Lausanne, Switzerland (June 9-11, 2002).
\bibitem{MOB03}
  MobiHoc 2003,
  Proc.\ of 4th ACM Int.\ Symp.\ on Mobile Ad Hoc Networking and
  Computing, Annapolis, MD, USA (June 1-3, 2003).
\bibitem{PRO95}
  J.G.\ Proakis, 
  {\em Digital Communications},
  McGraw-Hill, Singapore (1995).
\bibitem{PRA98}
  R.\ Prasad,
  {\em Universal Wireless Personal Communications},
  Artech House, Boston (1998).
\bibitem{RAP99}
  T.S.\ Rappaport, 
  {\em Wireless Communications -- Principles \& Practice},
  Prentice Hall, Upper Saddle River (1999).
\bibitem{ROD99}
  V.\ Rodoplu and T.H.\ Meng,
  IEEE Journal on Selected Areas in Communications -- 
  Special Issue on Ad Hoc Networks 17 (1999) 1333. 
\bibitem{GLA03}
  I.\ Glauche, W.\ Krause, R.\ Sollacher and M.\ Greiner, 
  Physica A 325 (2003) 577.
\bibitem{GUP00}
  P.\ Gupta and P.R.\ Kumar, 
  IEEE Trans.\ Inf.\ Theory IT-46 (2000) 388.
\bibitem{BET02}
  C.\ Bettstetter,
  {\em On the minimum node degree and connectivity of a wireless
  multihop network},
  in Proc.\ MOBIHOC 2002, pp.\ 80-91,
  Lausanne, Switzerland (June 9-11, 2002).
\bibitem{DOU02}
  O.\ Dousse, P.\ Thiran and M.\ Hasler,
  presented at IEEE INFOCOM 2002, New York (June 23-27, 2002), 
  http://www.ieee-infocom.org/2002/papers/481.pdf.
\bibitem{XUE03}
  F.\ Xue and P.R.\ Kumar,
  {\em The number of neighbors needed for connectivity of wireless 
  networks}, 
  Wireless Networks 10 (2004) 169. 
\bibitem{PRE92}
  W.H.\ Press, S.A.\ Teukolsky, W.T.\ Vetterling, and B.P.\ Flannery,
  {\em Numerical Recipes},
  Cambridge University Press, Cambridge (1992).
\bibitem{ROZ02}
  A.F.\ Rozenfeld, R.\ Cohen, D.\ ben-Avraham and S.\ Havlin,
  Phys.\ Rev.\ Lett.\ 89 (2002) 218701.
\bibitem{BAR03}
  M.\ Barthelemy,
  Europhys.\ Lett.\ 63 (2003) 915. 
\bibitem{HER03}
  C.\ Herrmann, M.\ Barthelemy and P.\ Provero,
  Phys.\ Rev.\ E 68 (2003) 026128.
\bibitem{MAN03}
  S.S.\ Manna, G.\ Mukherjee and P.\ Sen,
  arXiv:cond-mat/0307137.
\bibitem{DIJ59}
  E.\ Dijkstra,
  Numer.\ Math.\ 1 (1959) 269.
\bibitem{KUH03}
  F.\ Kuhn, R.\ Wattenhofer and A.\ Zollinger,
  {\em Worst-case optimal and average-case efficient geometric
  ad hoc routing},
  in Proc.\ MOBIHOC 2003, pp.\ 267-278,
  Annapolis, MD, USA (June 1-3, 2003).
\bibitem{WAT98}
  D.\ Watts and S.\ Strogatz,
  Nature 393 (1998) 440.
\bibitem{NEW01}
  M.E.J.\ Newman,
  Phys.\ Rev.\ E 64 (2001) 016132.

\end{thebibliography}
\end{document}